\documentclass[aps, prb, notitlepage, citeautoscript, superscriptaddress, eqsecnum, reprint]{revtex4-2}

\usepackage{xr}
\usepackage{amsmath}
\usepackage{amsfonts, amssymb,amsxtra}
\usepackage[]{graphicx}
\usepackage{grffile}
\usepackage{amsfonts}
\usepackage{framed}
\usepackage{bbm}
\usepackage{braket}
\usepackage{xcolor}
\usepackage{mathtools}
\usepackage{LatexCommands}

\usepackage[colorlinks=true]{hyperref}
\hypersetup{
    unicode=false,          
    pdftoolbar=true,        
    pdfmenubar=true,        
    pdffitwindow=false,     
    pdfstartview={FitH},    
    pdftitle={},    
    pdfauthor={},     
    pdfsubject={},   
    pdfcreator={},   
    pdfproducer={}, 
    pdfkeywords={} {} {}, 
    pdfnewwindow=true,      
    colorlinks=true,       
    linkcolor=magenta, 
    citecolor=blue,        
    filecolor=magenta,      
    urlcolor=blue           
}

\begin{document}
\title{Learning crystal field parameters using convolutional neural networks}

\author{Noah F.~Berthusen}
\affiliation{Ames Laboratory, Ames, Iowa 50011, USA}
\affiliation{Department of Electrical and Computer Engineering, Iowa State University, Ames, Iowa 50011, USA}
\author{Yuriy Sizyuk}
\affiliation{Ames Laboratory, Ames, Iowa 50011, USA}
\affiliation{Department of Physics and Astronomy, Iowa State University, Ames, Iowa 50011, USA}
\author{Mathias S.~Scheurer}
\affiliation{Department of Physics, Harvard University, Cambridge MA 02138, USA}
\affiliation{Institute for Theoretical Physics, University of Innsbruck, A-6020 Innsbruck, Austria}
\author{Peter P.~Orth}
\email{porth@iastate.edu}
\affiliation{Ames Laboratory, Ames, Iowa 50011, USA}
\affiliation{Department of Physics and Astronomy, Iowa State University, Ames, Iowa 50011, USA}

\begin{abstract}
    We present a deep machine learning algorithm to extract crystal field (CF) Stevens parameters from thermodynamic data of rare-earth magnetic materials. The algorithm employs a two-dimensional convolutional neural network (CNN) that is trained on magnetization, magnetic susceptibility and specific heat data that is calculated theoretically within the single-ion approximation and further processed using a standard wavelet transformation. We apply the method to crystal fields of cubic, hexagonal and tetragonal symmetry and for both integer and half-integer total angular momentum values $J$ of the ground state multiplet. We evaluate its performance on both theoretically generated synthetic and previously published experimental data on CeAgSb$_2$, PrAgSb$_2$ and PrMg$_2$Cu$_9$, and find that it can reliably and accurately extract the CF parameters for all site symmetries and values of $J$ considered. This demonstrates that CNNs provide an unbiased approach to extracting CF parameters that avoids tedious multi-parameter fitting procedures.
\end{abstract}
\date{\today}
\maketitle

\section{Introduction} 
\label{sec:introduction}
Rare-earth magnets often exhibit rich magnetic behaviors as a result of various competing energy scales that include spin-orbit coupling, crystal field (CF) and Zeeman energies as well as magnetic exchange interactions~\cite{elliottMagneticPropertiesRare1972,fuldeMagneticExcitationsCrystalfield1985,szytulaChapterMagneticProperties1991,canfieldPreservedEntropyFragile2016}. CFs play an important role as they cause magnetocrystalline anisotropies and in many cases determine the level degeneracies of the localized $f$ electron states~\cite{stevensMatrixElementsOperator1952,bleaneyParamagneticResonance1953,altshulerElectronParamagneticResonance1964,wybourneSpectroscopicPropertiesRare1965,kuzminChapterThreeTheory2007}. This strongly influences thermodynamic observables such as the magnetization, magnetic susceptibility, and specific heat~\cite{fazekasLectureNotesElectron1999}, but it can also have important ramifications for the nature of the Kondo effect in the system~\cite{coxQuadrupolarKondoEffect1987,coxExoticKondoEffects1998,Levy-PRL-1989,ikedaTheoryAnisotropicSemiconductor1996,Anders-PRL-2006,Peyker_2009,Dzero-PRL-2010,Romero_2013,Desgranges-Physica_B-2014, chandraHastaticOrderHeavyfermion2013,canfieldPreservedEntropyFragile2016,vandykeFieldinducedFerrohastaticPhase2019}.

CFs arise from time-reversal-even interactions between electrons (in $f$ orbitals for rare-earth elements) and charges in their crystalline environment and are conveniently described by an effective electrostatic potential. The form of the CF potential is dictated by the point symmetry at the rare-earth site and contains a variable number of independent parameters~\cite{betheTermaufspaltungKristallen1929,bleaneyParamagneticResonance1953,kuzminChapterThreeTheory2007}. For example, while the CF potential for $f$ electrons is fully described by only two independent parameters for the cubic point groups $\mathcal{G} = \text{m}\bar{3}\text{m}, 432, \bar{4}3\text{m}$, there are $26$ independent parameters for the lowest symmetry groups $1$ and $\bar{1}$~\cite{leaRaisingAngularMomentum1962,walterTreatingCrystalField1984}. These CF parameters are notoriously difficult to determine in first-principle calculations~\cite{richterChapterDensityFunctional2001}, and are therefore best regarded as phenomenological parameters that are found from comparison to experimental results. While most accurate values of CF parameters are obtained from analyzing inelastic neutron scattering results~\cite{loewenhauptChapterNeutronScattering1993,mozeChapterCrystalField1998}, much insight can already be gained by much more straightforward measurements of thermodynamic observables such as the (magnetic part of the) specific heat $c_M(T)$ as a function of temperature $T$, the magnetic susceptibility $\chi_a(T)$ along direction $a$, and the magnetization $\mu_a(\bfbb, T)$ in a finite magnetic field $\bfbb$. This approach allows investigating whole series of rare-earth compounds, which often provides a more complete understanding of a material class, as was demonstrated, for example, in Refs.~\cite{MYERS199927,budkoAnisotropyMetamagnetismRNi2Ge21999,PhysRevB.94.144434}.

Here, we focus on the method of extracting CF parameters from thermodynamic measurements that are performed in a regime above possible Kondo and magnetic ordering temperatures, where the rare-earth ion can be treated within the single-ion approximation~\cite{wybourneSpectroscopicPropertiesRare1965,kuzminChapterThreeTheory2007}. We will also assume that the Russell-Saunders approximation is valid and spin-orbit coupling is stronger than CF, Zeeman and magnetic exchange energy scales: $E_{\text{Coulomb}} \gg E_{\text{SOC}} \gg E_{\text{CF}}, E_{\text{Zeeman}}, E_{\text{ex}}$. Note that we will further focus on the case where the CF and Zeeman energies are larger than the exchange energy: $E_{\text{CF}}, E_{\text{Zeeman}} \gg E_{\text{ex}}$. Here, $E_{\text{Coulomb}}$ and $E_{\text{SOC}}$ refer to the isotropic Coulomb and spin-orbit interaction between $N$ electrons within the $4f^N$ electronic configuration of a single rare-earth ion, and $E_{\text{Zeeman}} = -\mu_B (\bfll + 2 \bfss) \cdot \bfbb$ with total orbital and spin angular momentum operators $\bfll$ and $\bfss$. Under these assumptions, one can restrict the attention to the ground state $J$ multiplet of the $4f^N$ configuration that is derived from the three Hund's rules~\cite{fazekasLectureNotesElectron1999}. Its $2J+1$ sub-levels are only degenerate for spherical symmetry and split in a crystalline environment into a sequence of lower order multiplets. While their multiplicity is fully determined by site symmetry, the energies of the different levels as well as their wave functions depend in general on the values of the CF parameters.

To obtain the CF parameters from measurements of thermodynamic observables, one traditionally proceeds as follows.  Starting from an initial guess of the CF parameters, one determines the energy levels and wave functions by diagonalizing the CF Hamiltonian $H_{\text{CF}} = \sum_{q, k} \mathcal{B}^q_k \widetilde{C}^{(k)}_q(\bfjj)$. Here, the summation runs over a symmetry-allowed set of quantum numbers $k$ and $q$ with $0 \leq k \leq 2 \ell, -k \leq q \leq k$ for a single-ion with orbital quantum number $\ell$ ($\ell = 3$ for $f$-electrons). The coefficients $\mathcal{B}^q_k$ are CF Stevens parameters and the CF operator ``equivalents'' $\widetilde{C}^{(k)}_q$ are expressed in terms of angular momentum operators $\bfjj$ acting on the ground state $J$ multiplet of the ion~\cite{stevensMatrixElementsOperator1952,wybourneSpectroscopicPropertiesRare1965,kuzminChapterThreeTheory2007,gtpack1,gtpack2}. Various forms for the operators, which differ in their normalization convention, have been used in the literature and will be discussed below. Once the energies and wave functions are known, it is straightforward to calculate thermodynamic observables such as $c_M, \chi_a$ and $\mu_a$ from the partition function in finite magnetic field (details are shown below). The theoretical result is then compared to experiment and the complete procedure is iterated with updated CF parameters until sufficient agreement is reached.

While this iterative process is straightforward in principle, it can be tedious and time consuming in practice, in particular for lower than cubic symmetries, where several fit parameters need to be optimized simultaneously.
This is complicated by the fact that the impact on the thermodynamic response that is caused by modifying the CF parameters $\{\mathcal{B}^q_k\}$ is in most cases unknown and not straightforward to derive.
This is a typical example of an ``inverse problem'' \cite{engl1996regularization} that often occurs in science, where one wants to estimate parameters $p$ characterizing the system (here the CF parameters) based on observations $O$ (in our case thermodynamic observables). Given a model $P$ (for us, the crystal-field Hamiltonian), it is straightforward to derive observables $O=F_P(p)$, but the inverse mapping $p=F_P^{-1}(O)$ is difficult to perform, in particular when the relation is non-linear as in our case; often, the inverse mapping is ill-conditioned or unstable and, thus, requires regularization.

Motivated by the multitude of recent explorations of machine-learning (ML) techniques in physics \cite{RMPMachineLearning,MEHTA20191,Dunjko_2018}, in general, and the success of artificial neural networks and other ML approaches to attack complex inverse problems of physics \cite{Arsenault_2017,PhysRevLett.124.056401,SupervisedAutoencoder,InverseDesignLaser,2019arXiv190911150L}, in particular, we here study how ML can be used to extract Stevens CF parameters from thermodynamic measurements.
This data-driven approach to inverse problems is based on first computing a large set of training data $\{(p_j,F_P(p_j))|j=1,2,\dots \}$, which requires solving the (simple) forward problem for many values of $p=p_j$. With this data set, a non-linear function is trained to reconstruct $p_j$ from $O_j=F_P(p_j)$; the key challenge is to find a model that generalizes well for feasible training data sizes, i.e., that works on physically relevant samples that are not part of the original training set.

More specifically, we here employ a convolutional neural network (CNN) to parametrize the non-linear function performing the inverse operation: it relates thermodynamic observables, $O=\{c_M(T), \chi_a(T), \mu_a(\bfbb, T)\}$, to a set of CF parameters $p=\{\mathcal{B}^q_k\}$. We train the CNN on thermodynamic data for different site symmetries (cubic $\text{m}\bar{3}\text{m}$, hexagonal $\bar{6}\text{m}2$, tetragonal $4\text{mm}$) and different values of angular momentum $J = 4$ and $J = 15/2$. This corresponds to the rare-earth ions Pr$^{3+}$ ($J=4$) and Er$^{3+}$ ($J = 15/2$) in different crystalline environments. The training data is obtained within the single-ion approximation, and further processed using a standard wavelet transformation before being fed into the CNN. We test the performance of the CNN on both calculated and previously published experimental data on CeAgSb$_2$~\cite{MYERS199927, Takeuchi-PRB-2003}, PrAgSb$_2$~\cite{MYERS199927} and PrMg$_2$Cu$_9$~\cite{PhysRevB.94.144434}. We find that our CNN architecture generalizes well for moderately large training data sets and for all site symmetries and values of $J$ considered. It also provides good estimates of the Stevens parameters from experimental data.

The remainder of the paper is organized as follows. In \secref{sec:crystal_field_thermodynamics}, we review the single-ion approximation, define our notation of the Stevens CF parameters, and explain how the relevant thermodynamic observables are computed. Readers already familiar with this, can proceed directly to \secref{sec:convolutional_neural_network_approach_to_finding_cf_parameters}, where we detail our proposed ML framework to estimate Stevens parameters from thermodynamic quantities. In \secref{sec:results} and \secref{sec:application_to_experimental_data}, we demonstrate and test our ML approach on synthetic and experimental data, respectively, and \secref{sec:summary_and_outlook} provides a summary.


\section{Crystal field thermodynamics in rare-earths} 
\label{sec:crystal_field_thermodynamics}
In this section, we provide
the necessary background to perform a quantitative analysis of CF effects on thermodynamic observables in rare-earth materials. We begin by describing the single-ion approximation, which assumes that interactions between different rare-earth ions are negligible. This approximation is often justified by the hierarchy of interactions that exist in rare-earth intermetallics~\cite{kuzminChapterThreeTheory2007}.
Focusing on the ground state multiplet of a single-ion with a definite total angular momentum $J$, we show how to expand the CF Hamiltonian for a given $J$ and point symmetry group $\mathcal{G}$ in terms of operator equivalents, as first introduced by Stevens \cite{stevensMatrixElementsOperator1952}.

Straightforward diagonalization of the Hamiltonian matrix together with elementary statistical mechanics calculations, then yield the thermodynamic observables, (i) specific heat $c_M$, (ii) magnetic susceptibilty $\chi_a$ (along direction $a$), and (iii) magnetization $\mu_a$ in finite applied magnetic field $B_a$. This calculation explicitly shows the (forward) mapping from a set of CF parameters to thermodynamic observables. These thermodynamic observables are then fed into the input nodes of a CNN that ``learns'' the inverse mapping from the observables to the CF parameters as output.

\subsection{Single-ion approximation} 
\label{sub:single_ion_approximation_and_operator_equivalents}
In the single-ion approximation one neglects the interaction between different rare-earth ions, which is often justified because the $4f$ electrons are strongly localized.
This leads to a relative weakness of $4f$-$4f$ exchange interactions compared to $3d$-$3d$ and $3d$-$4f$ interactions~\cite{kuzminChapterThreeTheory2007}, and an often weak hybridization between the localized $4f$ electrons and delocalized conduction electrons. The single-ion description breaks down, for example, when Kondo or Rudermann-Kittel-Kasuya-Yosida (RKKY) interactions play an important role in the magnetism of the system. Our analysis in the following is therefore restricted to parameter regimes, where both Kondo and RKKY interactions are weak effects, which is typically the case at not too low temperatures $T \gg T_{\text{K}}, T_{\text{RKKY}}$, where $T_{\text{K}}$ ($T_{\text{RKKY}}$) refer to Kondo and RKKY temperatures scales.

In the single-ion approximation, one describes the $4f$ electronic part of the system by a non-interacting collection of Hamiltonians for single rare-earth ions in a $4f^N$ configuration which each take the form~\cite{wybourneSpectroscopicPropertiesRare1965,kuzminChapterThreeTheory2007}
\begin{align}
    H_{4f} &= H_{\text{Coulomb}} + H_{\text{SOC}} - \mu_B (\bfll + 2 \bfss) \cdot \bfbb \nonumber \\
    & + \sum_{i=1}^N V_{\text{CF}}(r_i, \theta_i, \phi_i) \,.
\label{eq:2.1}
\end{align}
Here, $H_{\text{Coulomb}}$ and $H_{\text{SOC}}$ describe the isotropic Coulomb and spin-orbit interactions among the $N$ $4f$ electrons, which are the dominant energy scales. They enforce the three Hund's rules in the $4f^N$ configuration of the rare-earth ion, $S = \frac12 (2 \ell + 1 - |2 \ell + 1 -N|)$, $L = S(2 \ell + 1 - 2S)$, and $J = L \pm S$. The resulting ground state is then a $2J+1$ degenerate multiplet. Here, $\ell = 3$ is the orbital angular momentum of a single $f$ electron, $S$ ($L$) are the total spin (orbital) angular momentum quantum numbers and $J$ is the total angular momentum quantum number. The third Hund's rule enforces $J= L+S$ for more than half-filled $4f$ shells, $N \geq 2 \ell + 1$~\cite{wybourneSpectroscopicPropertiesRare1965,fazekasLectureNotesElectron1999}.

The third term in Eq.~\eqref{eq:2.1} describes the Zeeman coupling to an external magnetic field $\bfbb$,
where $\mu_B$ is the Bohr magneton and $\bfll = \sum_{i=1}^N \bfl_i$ and $\bfss = \sum_{i=1}^N \bfs_i$ denote total orbital and spin angular momenta of the $N$ electrons in the $4f^N$ configuration. In the following, we will assume that spin-orbit coupling dominates over Zeeman energy and use the Russell-Saunders LS-coupling scheme to express the Zeeman Hamiltonian using the total angular momentum $\bfjj = \bfll + \bfss$ as
\begin{align}
 H_{\text{Zeeman}} &= - \mu_B g_{JLS} \bfjj \cdot \bfbb \,.
\label{eq:2.1a}
\end{align}
Here, we have introduced the $g$-factor
\begin{align}
g_{JLS} &= 1 + \frac{J(J+1) + S(S+1) - L (L+1)}{2 J (J+1)} \,.
\label{eq:2.1b}
\end{align}
with angular momentum quantum numbers $J, L, S$ corresponding to the magnitude of the operators $\bfjj, \bfll, \bfss$, respectively.

Finally, the last term in Eq.~\eqref{eq:2.1} denotes the CF potential, which can be expanded in a series of (single-particle) irreducible tensor operators as~\cite{wybourneSpectroscopicPropertiesRare1965}
\begin{align}
 V_{\text{CF}}(r, \theta, \phi) &= \sum_{k=2, 4, 6} \sum_{q = -k}^k B^q_k(r) C^{(k)}_{q}(\theta, \phi)\,.
\label{eq:2.2}
\end{align}
Here, the functions $B^q_k(r)$ depend on the radial coordinate only, and $C^{(k)}_{q}(\theta, \phi) = \sqrt{\frac{4 \pi}{2 k + 1}} Y^q_k(\theta, \phi)$ are related to the spherical harmonics.
Both sets of operators, $B^q_k(r)$ and $C^{(k)}_q(\theta, \phi)$, act on the coordinates $\bfr_i$ of individual electrons in the $f$-shell. Note that the summation of $k$ is restricted to $k = 2,4, 6$, as we anticipate to evaluate matrix elements of $V_{\text{CF}}$ only within a single $4f^N$ configuration. This excludes odd values of $k$ by parity considerations. Higher values of $k > 6$ are excluded from the triangular condition $k \leq 2 \ell $ of the Clebsch-Gordon coefficients (or Wigner $3j$ symbols), which arise when performing an integration over products of three spherical harmonics~\cite{edmondsAngularMomentumQuantum1957}. Finally, we have also excluded the $k=0$ term as it amounts to an unimportant constant energy shift.


\subsection{Operator equivalents in crystal field Hamiltonians} 
\label{sub:operator_equivalents}
The evaluation of matrix elements of the CF Hamiltonian
\begin{align}
H_{\text{CF}} = \sum_{i=1}^N \sum_{k= 2, 4, 6} \sum_{q = -k}^k B^q_k(r_i) C^{(k)}_q(\theta_i, \phi_i)
\label{eq:2.3}
\end{align}
in the limited subspace of a $4f^N$ electronic configuration of a single rare-earth ion is made easier by the method of operator equivalents introduced by Stevens~\cite{stevensMatrixElementsOperator1952}. First, within a fixed $4f^N$ manifold, the radial operators can be replaced by their expectation values in the $4f$ states, which defines the (single-particle) Stevens coefficients $B^q_k \equiv \av{B^q_k(r_i)}_{4f}$. Since the precise form of the wavefunction is difficult to determine, a theoretical calculation of the Stevens coefficients from first-principles is notoriously challenging~\cite{richterChapterDensityFunctional2001}. The $B^q_k$ are therefore best regarded as phenomenological coefficients that are obtained from a comparison of calculated physical observables to experimental data.

The method of operator equivalents~~\cite{stevensMatrixElementsOperator1952,juddOperatorTechniquesAtomic1963,altshulerElectronParamagneticResonance1964,buckmasterTablesMatrixElements1962, smithUseOperatorEquivalents1966} relates matrix elements of (the sum over) irreducible tensor operators within a $4f^N$ configuration to matrix elements of expressions that depend on angular momentum operators $\bfl_i$:
\begin{multline}
    \braket{\{l_i,m_i\} | \sum_{i=1}^N C^{(k)}_q(\theta_i, \phi_i) | \{l_i, m'_i\}} \\
    = \mathfrak{a}_k \braket{\{l_i,m_i\} | \sum_{i=1}^N \widetilde{C}^{(k)}_q(\bfl_i) | \{l_i, m'_i\}} \,.
\label{eq:2.4}
\end{multline}
Here, $\mathfrak{a}_k$ is an $k$ (and $l_i$) dependent coefficient. The operator expressions on both sides transform under rotation according to the same irreducible representation of the continuous rotation group. This condition in fact defines the ``operator equivalent'' of the irreducible tensor operator on the left-hand side. The operator equivalents $\widetilde{C}^{(k)}_q(\bfl_i)$ can be obtained by converting the functions $C^{(k)}_q(\theta, \phi)$ into Cartesian coordinates, $(x,y,z) = (\sin \theta \sin \phi, \sin \theta \cos \phi,\cos \theta)$, symmetrizing monomials (e.g., $xy \rightarrow (xy+yx)/2$), and replacing $(x_i/r_i, y_i/r_i, z_i/r_i) \rightarrow (l_x, l_y, l_z)_i$.
The proportionality of the matrix elements in Eq.~\eqref{eq:2.4} relies on the fact that the rotation group is continuous and, loosely speaking, matrix elements for any point on the sphere can thus be obtained from those at a single, fixed point via rotation. The proportionality factors $\mathfrak{a}_k$ essentially account for the difference of the matrix elements at the single reference point. The $\mathfrak{a}_k$ are independent of $q$ due to the Wigner-Eckart theorem~\cite{edmondsAngularMomentumQuantum1957}. In the literature, the proportionality coefficients are typically denoted as $\mathfrak{a}_2 = \alpha_l$, $\mathfrak{a}_4 = \beta_l$, and $\mathfrak{a}_6 = \gamma_l$. For $l_i = 3$ corresponding to $4f$ rare-earth ions, one finds the values $\mathfrak{a}_2 = -2/45$, $\mathfrak{a}_4 = 2/495$, and $\mathfrak{a}_6 = -4/3861$~\cite{bleaneyParamagneticResonance1953,kuzminChapterThreeTheory2007,gtpack2}.

Here, we restrict our analysis to the $(2J+1)$-dimensional ground state multiplet of the $4f^N$ configuration that obeys the three Hund's rules. Combining individual angular momenta to the total orbital angular momentum $\bfll = \sum_{i=1}^N \bfl_i$ and considering spin-orbit coupling within a fixed LS term, leading to total angular momentum $\bfjj = \bfll + \bfss$, one can derive a similar ``operator equivalent'' relation as Eq.~\eqref{eq:2.4} for matrix elements taken within a particular $J$ multiplet
\begin{multline}
\braket{L, S, J, M_J | \sum_{i=1}^N C^{(k)}_q(\theta_i, \phi_i) | L, S, J, M'_J} \\
= \mathfrak{b}_k \braket{L, S, J, M_J | \widetilde{C}^{(k)}_{q} (\bfjj)  | L, S, J, M'_J} \,.
\label{eq:2.5}
\end{multline}
Here, the coefficients $\mathfrak{b}_k$ depend on $k$ as well as on the quantum numbers $l_i, L, S, J$. Like the $\mathfrak{a}_k$, they are independent of $m$ due to the Wigner-Eckart theorem. The values of the $\mathfrak{b}_k$ for the ground state multiplets of the $R^{3+}$ rare-earth ions can be found in the literature, where they are commonly denoted as $\mathfrak{b}_2 = \alpha_J = \theta_2$, $\mathfrak{b}_4 = \beta_J = \theta_4$ and $\mathfrak{b}_6 = \gamma_J = \theta_6$~\cite{wybourneSpectroscopicPropertiesRare1965,kuzminChapterThreeTheory2007}.

Using the operator equivalence in Eq.~\eqref{eq:2.5}, one can thus express the CF Hamiltonian acting within the $(2J+1)$-dimensional ground state multiplet as
\begin{align}
 H_{\text{CF}} = \sum_{k = 2,4,6} \sum_{q= -k}^k \mathcal{B}^q_k \widetilde{C}^{(k)}_q(\bfjj)\,.
\label{eq:2.6}
\end{align}
Here, we have introduced the Stevens coefficients $\mathcal{B}^q_k =  \mathfrak{b}_k B^q_k$ that depend on the radial expectation values through $B^q_k \equiv \av{B^q_k(r_i)}_{\text{4f}}$ (see Eq.~\eqref{eq:2.2}). We regard both $B^q_k$  and $\mathcal{B}^q_k$ as phenomenological parameters that are determined by comparing theoretical calculations of physical observables to experimental results.

In the following, we will use the CF Hamiltonian of the form in Eq.~\eqref{eq:2.6}.
We note that in the literature it is common to use the so-called Stevens operator equivalents $O^q_k(\bfjj)$~\cite{leaRaisingAngularMomentum1962,danielsenQuantumMechanicalOperator1972,kuzminChapterThreeTheory2007,gtpack2}, which are based on the tesseral harmonics (real and imaginary parts of the spherical harmonics). We denote the corresponding Stevens coefficients multiplying $O^q_k$ as $B^q_{k,\text{Stevens}}$ in the following. The Stevens operators employ a different normalization convention than the irreducible tensor operator equivalents $\widetilde{C}^{(k)}_q(\bfjj)$. This requires using $k$ and $q$ dependent factors $K^q_k$ relating the $\widetilde{C}^{(k)}_q$ and $O^q_k$ operators: $O^q_k = \frac{1}{K^q_k} \frac{2 k + 1}{4 \pi} \frac{1}{\sqrt{2}} \bigl[\widetilde{C}^{(k)}_{-q} + (-1)^q \widetilde{C}^{(k)}_q \bigr]$ for $q \neq 0$ and $O^0_k = \frac{1}{K^0_k}\frac{2 k + 1}{4 \pi} \widetilde{C}^{(k)}_q$~\cite{danielsenQuantumMechanicalOperator1972}. The factors $K^q_k$ can be found, for example, in Ref.~\cite{danielsenQuantumMechanicalOperator1972}, but can also be easily derived by direct comparison of the operator matrices~\cite{gtpack1,gtpack2}. The main disadvantage of the Stevens operators $O^q_k(\bfjj)$ is that they do \emph{not} obey the Wigner-Eckart theorem. Their matrix elements are explicitly tabulated~\cite{abragamElectronParamagneticResonance1970,buckmasterTablesMatrixElements1962,gtpack2}.

\subsection{Stevens crystal field parameters} 
\label{sub:stevens_parameters}
In this section, we describe the convention of Stevens parameters that we use in the following and their relation to other common definitions. Following Lea, Leask, Wolf~\cite{leaRaisingAngularMomentum1962} and Walter~\cite{walterTreatingCrystalField1984}, it is convenient to perform a transformation from the Stevens parameters $\{\mathcal{B}^q_k\}$ in Eq.~\eqref{eq:2.6} to a set of Stevens coefficients $\{x_0, \ldots, x_{N_{\text{St}}-1}\}$. Here, $x_0$ describes the overall energy scale of the CF splitting (note that $x_0$ can be negative). The dimensionless parameters $\{x_1, \ldots, x_{N_{\text{St}}-1}\}$ fulfill $|x_i| \leq 1$ and describe the relative weight of the different Stevens parameters $\mathcal{B}^q_k$.

\subsubsection{Cubic symmetry} 
\label{ssub:cubic_symmetry}
Let us explicitly describe the transformation from $\{ \mathcal{B}^q_k\} \rightarrow \{x_i\}$ for cubic symmetry. The derivation easily generalizes to arbitrary point groups $\mathcal{G}$. For the cubic point groups $\mathcal{G} = \{ \text{m$\bar{3}$m}, \text{432}, \bar{4}3\text{m} \}$, the CF Hamiltonian contains two independent Stevens parameters, $N_{\text{St}} = 2$, and reads
\begin{align}
H_{\text{CF}} &= \mathcal{B}^4_4 \Bigl( \widetilde{C}^{(4)}_4 + \widetilde{C}^{(4)}_{-4} + \sqrt{\frac{14}{5}} \widetilde{C}^{(4)}_0 \Bigr) \nonumber \\ & \quad + \mathcal{B}^4_6 \Bigl( \widetilde{C}^{(6)}_4 + \widetilde{C}^{(6)}_{-4} - \sqrt{\frac{2}{7}} \widetilde{C}^{(6)}_0 \Bigr)\,.
\label{eq:2.8}
\end{align}
Let us first normalize each operator that multiplies a particular Stevens coefficient
\begin{align}
\widetilde{\mathcal{O}}^{(4)} = \widetilde{C}^{(4)}_4 + \widetilde{C}^{(4)}_{-4} + \sqrt{\frac{14}{5}} \widetilde{C}^{(4)}_0
\label{eq:2.8a}
\\
\widetilde{\mathcal{O}}^{(6)} = \widetilde{C}^{(6)}_4 + \widetilde{C}^{(6)}_{-4} - \sqrt{\frac{2}{7}} \widetilde{C}^{(6)}_0
\label{eq.2.8b}
\end{align}
Normalization of $\widetilde{\mathcal{O}}^{(4)}$ and $\widetilde{\mathcal{O}}^{(6)}$ can be achieved by dividing by the sum of squared eigenvalues $\Lambda^{(k)} = \sqrt{\sum_{i=1}^{2J+1} |\lambda^{(k)}_i}|^2$, where $\lambda^{(k)}_i$ are the eigenvalues of the operator $\widetilde{\mathcal{O}}^{(k)}$. Specifically, we define the scaled operators
\begin{align}
O^{(k)} = \frac{2J+1}{\Lambda^{(k)}} \widetilde{\mathcal{O}}^{(k)}
\label{eq:2.9}
\end{align}
and express the CF Hamiltonian as
\begin{align}
H_{\text{CF}} &= \mathcal{B}^4_4 \widetilde{\mathcal{O}}^{(4)} + \mathcal{B}^4_6 \widetilde{\mathcal{O}}^{(6)} \nonumber \\
&= x_0 \Bigl[x_1 \mathcal{O}^{(4)} + (|x_1| - 1) \mathcal{O}^{(6)} \Bigr]\,.
\label{eq:2.10}
\end{align}
As anticipated above, the scale parameter $x_0$, which can be positive or negative, sets the overall energy scale of the CF splitting. The dimensionless weight parameter $x_1$ describes the ratio of Stevens coefficients
\begin{align}
\frac{\mathcal{B}^4_4}{\mathcal{B}^4_6} \propto \frac{x_1}{|x_1| - 1}\,,
\label{eq:2.10a}
\end{align}
and lies in the interval $-1 \leq x_1 \leq 1$. The ratio $\mathcal{B}^4_4/\mathcal{B}^4_6 = 0$ corresponds to $x_1 = 0$, whereas $\mathcal{B}^4_4/\mathcal{B}^4_6 \rightarrow \pm \infty$ corresponds to $x_1 = \pm 1$. The exact relation between the two sets of Stevens parameters $\{ \mathcal{B}^q_k \}$ and $\{x_i\}$ depends on the value of $J$ and can b easily rederived from Eq.~\eqref{eq:2.10}.

\subsubsection{Hexagonal symmetry} 
\label{ssub:hexagonal_symmetry}
For hexagonal site symmetry with point symmetry groups $\mathcal{G} = \{\bar{6}\text{m}2, 6/\text{mmm}, 6\text{mm}, 622\}$, the CF Hamiltonian contains four independent Stevens parameters, $N_{\text{St}} = 4$, and reads
\begin{align}
H_{\text{CF}} &= \mathcal{B}^0_2 \widetilde{C}^{(2)}_0 + \mathcal{B}^0_4 \widetilde{C}^{(4)}_0 + \mathcal{B}^0_6 \widetilde{C}^{(6)}_0 + \mathcal{B}^6_6 \bigl( \widetilde{C}^{(6)}_6 + \widetilde{C}^{(6)}_{-6} \bigr)\,.
\label{eq:2.11}
\end{align}
Defining $\widetilde{\mathcal{O}}^{(k)}_q = \widetilde{C}^{(k)}_q$ for $ m \neq 6$ and $\widetilde{\mathcal{O}}^{(6)}_6 = \widetilde{C}^{(6)}_6 + \widetilde{C}^{(6)}_{-6}$ for $l=m=6$, we again normalize the $\widetilde{\mathcal{O}}^{(k)}_q$ via
\begin{subequations}\begin{align}
\mathcal{O}^{(k)}_q &= \frac{2J+1}{\Lambda^{(k)}_q} \widetilde{\mathcal{O}}^{(k)}_q\,.
\label{eq:2.12}
\end{align}
Here,
\begin{align}
\Lambda^{(k)}_q = \sqrt{\sum_{i=1}^{2J+1} |\lambda^{(k)}_{q,i}|^2}\,,
\label{eq:2.12a}
\end{align}
\label{GeneralNormalization}\end{subequations}
where $\lambda^{(k)}_{q,i}$ are the eigenvalues of $\widetilde{\mathcal{O}}^{(k)}_q$. Finally, we express the CF Hamiltonian in terms of the normalized operators as
\begin{align}
H_{\text{CF}} & = |x_0| \Bigl[ x_1 \mathcal{O}^{(2)}_0 + x_2 \mathcal{O}^{(4)}_0 + x_3 \mathcal{O}^{(6)}_6 \nonumber \\
& \quad + \text{sign}(x_4)\Bigl|1 - |x_1| - |x_2| - |x_3|\Bigr| \mathcal{O}^{(6)}_0 \Bigr] \,.
\label{eq:2.13}
\end{align}
As before, $x_0$ describes the overall energy scale, whereas the weight parameters $-1 \leq x_1, \ldots x_3 \leq 1$ describe the relative weight of the four Stevens parameters $\mathcal{B}^q_k$. Note that we have split off the sign of $x_0$. This turns out to be advantageous in the ML calculation described below as it makes the overall scale prefactor $|x_0|$ strictly positive. This comes at the cost of introducing an additional parameter, $x_4$, defined as $\text{sign}(x_4) \coloneqq \text{sign}(x_0)$. Only the sign of $x_4$ enters the Hamiltonian.

\subsubsection{Tetragonal symmetry} 
\label{ssub:tetragonal_symmetry}
For tetragonal site symmetry with point symmetry groups $\mathcal{G} = \{4\text{mm}, 4/\text{mmm}\}$, the CF Hamiltonian contains five independent Stevens parameters, $N_{\text{St}} = 5$, and reads
\begin{align}
H_{\text{CF}} &= \mathcal{B}^0_2 \widetilde{C}^{(2)}_0 + \mathcal{B}^0_4 \widetilde{C}^{(4)}_0 + \mathcal{B}^4_4 \bigl( \widetilde{C}^{(4)}_{4} + \widetilde{C}^{(4)}_{-4} \bigr) + \mathcal{B}^0_6 \widetilde{C}^{(6)}_0 \nonumber \\
& \quad + \mathcal{B}^4_6 \bigl( \widetilde{C}^{(6)}_4 + \widetilde{C}^{(6)}_{-4} \bigr) \nonumber \\
&= \mathcal{B}^0_2 \widetilde{\mathcal{O}}^{(2)}_0 + \mathcal{B}^0_4 \widetilde{\mathcal{O}}^{(4)}_0 + \mathcal{B}^4_4 \widetilde{\mathcal{O}}^{(4)}_{4}+ \mathcal{B}^0_6 \widetilde{\mathcal{O}}^{(6)}_0 + \mathcal{B}^4_6 \widetilde{\mathcal{O}}^{(6)}_4\,.
\label{eq:2.18}
\end{align}
We have defined the operators $\widetilde{\mathcal{O}}^{(k)}_q$ in the second line, which we then normalize as in \equref{GeneralNormalization}.

Finally, the CF Hamiltonian is expressed in terms of the normalized operators as
\begin{align}
H_{\text{CF}} & = |x_0| \Bigl[ x_1 \mathcal{O}^{(2)}_0 + x_2 \mathcal{O}^{(4)}_4 + x_3 \mathcal{O}^{(4)}_0 + x_4 \mathcal{O}^{(6)}_4 \label{eq:2.20} \\
& \quad + \text{sign}(x_5) \Bigl|1 - |x_1| - |x_2| - |x_3| - |x_4| \Bigr| \mathcal{O}^{(6)}_0 \Bigr] \,.
\nonumber
\end{align}
In addition to the scale parameter $x_0$, the Hamiltonian contains four bounded Stevens parameters $-1\leq x_1, \ldots, x_4 \leq 1$. Like in the hexagonal case, we have split off the sign of $x_0$ explicitly and introduced an additional parameter, $x_5$, as $\text{sign}(x_5) \coloneqq \text{sign}(x_0)$. The Hamiltonian only depends on the sign of $x_5$.


\subsection{Thermodynamic observables} 
\label{sub:thermodynamic_observables}
In this section, we describe how to obtain the thermodynamic observables of interest: magnetization (per rare-earth ion), $\boldsymbol{\mu}(T, \bfbb)$, in finite magnetic field, magnetic susceptibility $\chi_a(T)$ along direction $a$, and specific heat $c_M(T)$. We calculate these quantities starting from the Hamiltonian (\ref{eq:2.1}) of a single rare-earth ion in a magnetic field $\bfbb$ and exposed to a CF with point symmetry $\mathcal{G}$. From our discussion above, we know that the Hamiltonian projected onto the ground state multiplet with total angular momentum $J$ reads
\begin{align}
 H_J &=  - \mu_B g_{JLS} \bfjj \cdot \bfbb + \sum_{l = 2,4,6} \sum_{m= -l}^l \mathcal{B}^q_k \, \widetilde{C}^{(k)}_q(\bfjj) \,.
\label{eq:2.22}{}
\end{align}
Here, $\bfjj = (J_x, J_y, J_z)$ denotes the total angular momentum operator, the $g$-factor $g_{JLS}$ is explicitly given in Eq.~\eqref{eq:2.1b}, and the form of the CF Hamiltonian is constrained by the point group $\mathcal{G}$. The method we describe in the following can be used for any point group $\mathcal{G}$, but we will focus on the experimentally common cases of cubic, hexagonal and tetragonal crystal symmetry with  point groups $\mathcal{G}$ that were discussed in Sec.~\ref{sub:stevens_parameters}.

\begin{figure*}[tb]
    \centering
    \includegraphics[width=\linewidth]{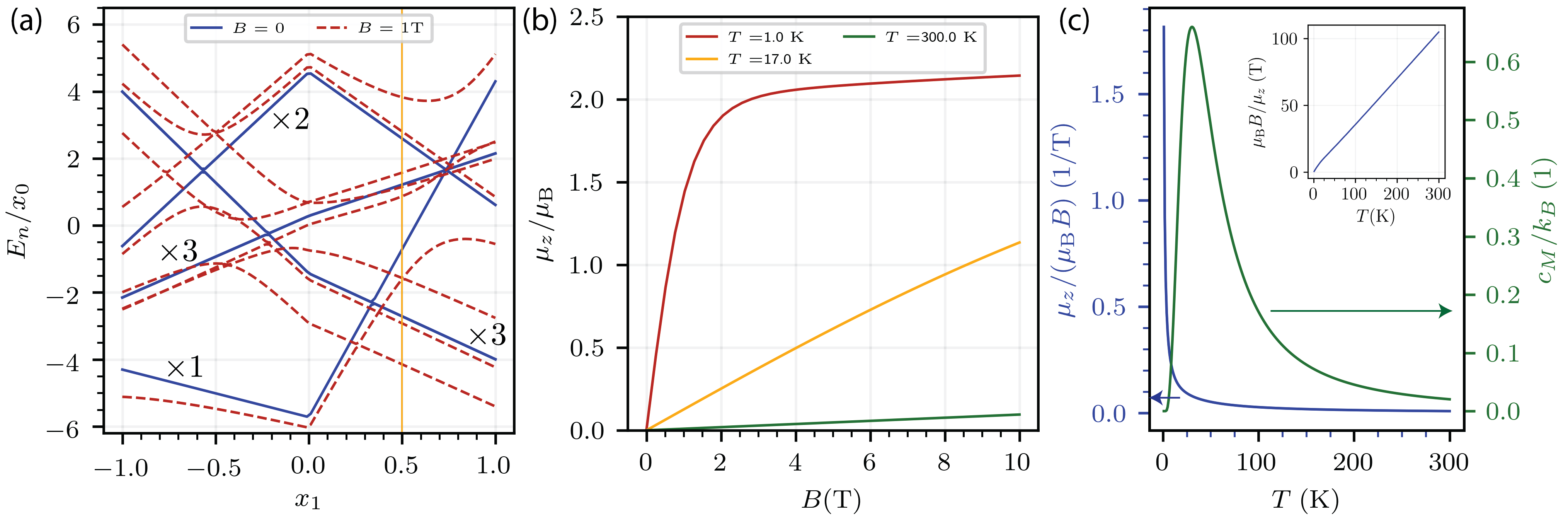}
    \caption{(a) Energy levels $E_n/x_0$ of a $J=4$ rare-earth ion (Pr$^{3+}$) in cubic CF with point symmetry $\mathcal{G} = \text{m}\bar{3}\text{m}$ as a function of Stevens parameter $x_1$. The degeneracy of the levels in the absence of a magnetic field is indicated in the figure, and splits in applied magnetic field $B$ along the $z$ axis. The $g$-factor is set to $g_{JLS} = 4/5$. The vertical yellow line indicates the parameter choice, $x_1 = 0.5$, for panels $(b, c)$, where we further set $x_0 = 20$~K as absolute energy scale. (b) Magnetization per rare-earth ion $\mu_z/\mu_B$ as a function of applied magnetic field $B$ along the $z$-axis for different temperatures. The Stevens parameters are $x_0 = 20$~K and $x_1 = 0.5$. The fully saturated moment is reduced from the value in the ground multiplet at $x_1 = 0.5$ is $\mu_z/\mu_B = 5/2$ due to field-induced mixing into other states. (c) Magnetic susceptibility along $z$-axis (per rare-earth ion) $\chi_z/\mu_B = \mu_z/(\mu_B B)$ (blue, left $y$-axis) and specific heat $c_M/k_B$ (green, right $y$-axis) as a function of temperature $T$. The magnetic field is fixed to $B = 10^{-4}$~T when computing $\chi_z$. The Stevens parameters are identical to panel (b). The inset shows the inverse susceptibility $\chi_z^{-1} = \mu_B B/\mu_z$, highlighting the Curie-like behavior that occurs over the full temperature range due to the triplet ground state. The specific heat $c_M/k_B$ shows a Schottky-like peak at a position proportional to the level splitting between the ground multiplet and the excited states $\Delta = 40$~K. Note that here the first excited state is a singlet and the contribution of the next higher triplet level (at about $80$~K) is significant. }
    \label{fig:training_data_set-J_4-Oh}
\end{figure*}

In the basis of $J_z$ eigenstates, $J_z \ket{m_J} = m_J \ket{m_J}$, the Hamiltonian $H_J$ is a $(2J+1)\times (2J+1)$ dimensional matrix that can be easily diagonalized,
\begin{align}
 H_J \ket{n} = E_n \ket{n},
\label{eq:2.23}
\end{align}
with field-dependent energies $E_n$ and eigenstates
\begin{align}
\ket{n} = \sum_{m_J=-J}^J a_{m_J} \ket{m_J} \,.
\label{eq:2.24}
\end{align}
In Fig.~\ref{fig:training_data_set-J_4-Oh}(a), we show the resulting energy spectrum for $J=4$ and cubic CF ($\mathcal{G} = \text{m}\bar{3}\text{m}$) as a function of the dimensionless Stevens parameter $x_1$, see \equref{eq:2.10}. The level spectrum is presented both at zero magnetic field and at a finite field of $B_z = 1$~T. In zero field, we observe that for $x_0 > 0$, the ground state is a singlet (triplet) for $x_1 < 0.34$ ($x_1 > 0.34$). For $x_0 < 0$, the ground state is a triplet for $x_1 < - 0.57$, a doublet for $-0.57 < x_1 < 0.74$ and a singlet for $x_1 > 0.74$. The level degeneracy is split in an applied magnetic field and one observes the emergence of several (avoided) level crossings. A similar behavior is observed for other integer values of the angular momentum quantum number $J$ with lower degeneracies in the case of lower-symmetry CFs. For half-integer $J$, the levels are at least doubly degenerate in the absence of an external magnetic field due to Kramers theorem.

\subsubsection{Magnetization and magnetic susceptibility} 
\label{ssub:magnetization}
The magnetization per single rare-earth ion along direction $a$ is given by
\begin{align}
 \mu_a(T, \bfbb) = \frac{\mu_B g_{JLS}}{Z} \sum_n \braket{n | J_a | n} e^{- E_n/k_B T}
\label{eq:2.25}
\end{align}
with partition function $Z = \text{Tr} e^{- \beta H_J}$ and magnetic field along direction $a$. In Fig.~\ref{fig:training_data_set-J_4-Oh}(b), we show the magnetization $\mu_z$ as a function of $\vec{B}=(0,0,B)^T$ for different temperatures $T$ in a cubic CF $(\mathcal{G} = \text{m}\bar{3} \text{m})$. The Stevens parameters are chosen to be $x_0 = 20$~K and $x_1 = 0.5$ such that the ground state is a triplet with $\av{J_z} = \pm \frac{5}{2}, 0$. The magnetization $\mu_z$ thus increases linearly at low fields with a slope that increases Curie-like as $1/T$. The magnetization saturates at a saturation magnetic field value $B_{\text{sat}}$ that increases with temperature. At the lowest temperature, $T = 1$~K, the saturation occurs at $B_{\text{sat}}(1~\text{K}) \approx 3$~T. The saturation value of the magnetization is given by $\mu_z^{\text{sat}} = \mu_B g_{JLS} \braket{0 | J_z |0}$, where $\ket{0}$ is the ground state in magnetic field. Here, we have chosen $L = 5$ and $S = 1$ such that $g_{JLS} = 4/5$, which corresponds to the rare-earth ion Pr$^{3+}$. Note that $\mu_z^{\text{sat}}$ deviates slightly from the expected value of $5/2 \times 4/5 = 2$, where $5/2$ is the expectation value of $J_z$ in the triplet ground state in small fields, due to field induced mixing of higher levels.

The magnetic susceptibility is obtained at small magnetic fields from the slope
\begin{align}
    \chi_{a}(T) = \mu_a(T,B_a)/B_a \,.
\label{eq:2.26}
\end{align}
Its behavior at low temperatures is determined by the ground state degeneracy~\cite{fazekasLectureNotesElectron1999}. If the ground state is a singlet, it is of van-Vleck type, $\chi_a \propto \sum_{i\neq 0} \frac{|\braket{i | J_a | 0}|^2}{E_{i} - E_0}$, and becomes temperature independent at temperatures much smaller than the energy gap to the first excited state, $k_B T \ll E_1 - E_0$. In contrast, $\chi_a$ is Curie-like $\chi_a \propto \frac{g_{JLS}^2 \mu_B^2 \braket{0 | J^2 | 0}}{T}$, if the ground state degeneracy is larger than one. In Fig.~\ref{fig:training_data_set-J_4-Oh}(c), we show the susceptibility $\chi_z$ as a function of temperature for the case of a triplet ground state, where it follows a characteristic Curie-like behavior $\chi_a \propto 1/T$ [see inset of Fig.~\ref{fig:training_data_set-J_4-Oh}(c)].

\subsubsection{Specific heat} 
\label{ssub:specific_heat}
The specific heat in zero magnetic field is calculated from
\begin{align}
c_M(T) &= \frac{1}{k_B T^2}\Bigl( \av{H_\text{CF}^2} - \av{H_{\text{CF}}}^2 \Bigr) \,,
\label{eq:2.27}
\end{align}
where the average is performed with respect to the CF eigenstates $H_{\text{CF}} \ket{n^{(0)}} = E^{(0)}_n \ket{n^{(0)}}$:
\begin{align}
\av{\mathcal{O}} = \frac{1}{Z_{\text{CF}}} \sum_{n} \langle n^{(0)} | \mathcal{O} | n^{(0)} \rangle e^{- E_{n}^{(0)}/k_B T}
\label{eq:2.28}
\end{align}
with $Z_{\text{CF}} = \text{Tr} e^{- H_{\text{CF}}/k_B T} = \sum_{n} e^{- E_{n}^{(0)}/k_B T}$. The specific heat vanishes exponentially, $c_M \propto \Bigl( \frac{\Delta}{k_B T}\Bigr)^2 e^{- \Delta/k_B T}$, at temperatures below the gap to the first excited state, $k_B T \ll \Delta$. As shown in Fig.~\ref{fig:training_data_set-J_4-Oh}(c) for $J=4$ and $\mathcal{G} = \text{m} \bar{3} \text{m}$, it exhibits a Schottky anomaly peak at higher temperatures, whose position and weight yields direct information about the size of the gap to the excited states and the relative degeneracies of the ground and excited state levels. Note that excited state levels higher than the first often occur nearby in energy and thus contribute to the specific heat as well.



\section{CNN approach for finding crystal field parameters} 
\label{sec:convolutional_neural_network_approach_to_finding_cf_parameters}
In this section, we describe the method of using a two-dimensional CNN to determine the Stevens parameters $\{x_i\}$ for a given angular momentum $J$ and CF symmetry group $\mathcal{G}$ from thermodynamic observables. Our goal is to build a ML model that can be fed with experimental data and accurately predict the underlying Stevens coefficients that characterize the material, thereby circumventing a time-consuming data fitting procedure. One therefore places thermodynamic data on the input nodes of the network and obtains the set of Stevens coefficients as output. We choose the input data of the network to be from observables that are experimentally readily available: magnetization $\mu_a(T, B_a)$, magnetic susceptibility $\chi_a(T)$, and magnetic specific heat $c_M(T)$. To train the network, we require a sufficiently large dataset that we generate by calculating the thermodynamic observables for random choices of Stevens parameters within the single-ion approximation as described in Sec.~\ref{sec:crystal_field_thermodynamics}. Comparing different network architectures, we found it to be advantageous to perform a wavelet transformation on the data before feeding it into the network. In the following, we describe the details of the training data generation and the network architecture and parameters.

\subsection{Training data generation} 
\label{sec:training_data_generation}
\subsubsection{Thermodyamic data generation}
A training data set contains the following three types of observables, which are calculated for a fixed choice of angular momentum $J$ and point group $\mathcal{G}$, and randomly sampled Stevens coefficients $\{x_i\}$, using the approach detailed in Sec.~\ref{sec:crystal_field_thermodynamics}.
\begin{figure}
    \centering
    \includegraphics[width=1\linewidth]{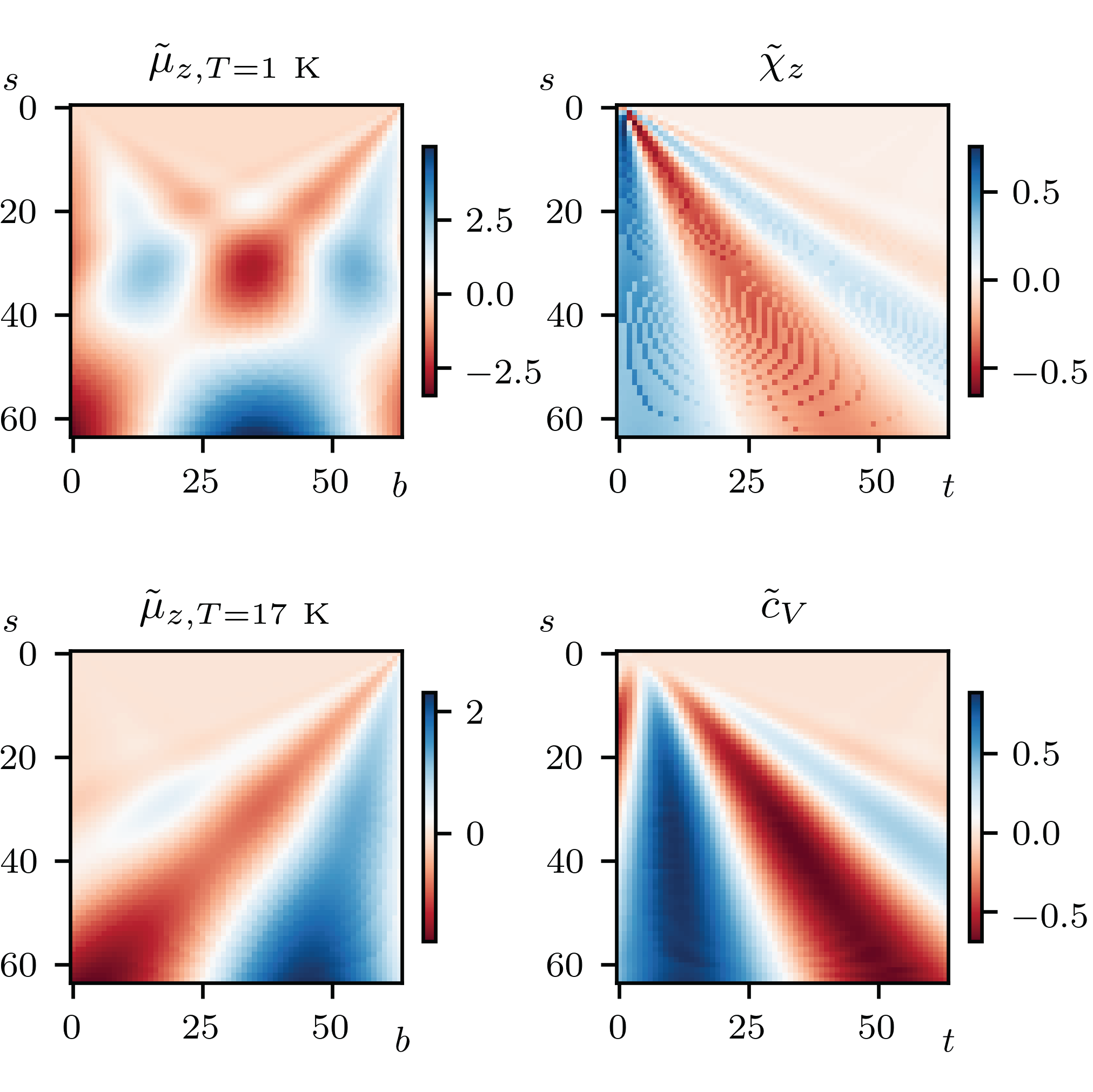}
    \caption{Continuous wavelet scaleograms of the data of Fig.~\ref{fig:training_data_set-J_4-Oh}(b,c) for which $J=4$, $\mathcal{G} = \text{m}\bar{3}\text{m}$, $x_0 = 20$~K, and $x_1 = 0.5$. The plots show the CWT coefficients, using the real Morlet mother wavelet (\ref{MamaMorlet}), as a function of scale $s$ and physical index $b$ (or $t$). We choose linearly spaced scales from $s_\text{min} = 1$ to $s_{\text{max}} = 64$, and also use 64 equally spaced points along the ``physical'' dimension, corresponding to the magnetic field, $b$ (see Eq.~\eqref{eq:3.1}), and the temperature, $t$ see Eq.~\eqref{eq:3.2}).
    In addition to these four scaleograms we also provide $\tilde{\mu}_{z,T = 300\,\text{K}}$ to the CNN, resulting in a total of five scaleograms to be layered into one training sample. For the lower symmetry point groups, we provide magnetization and susceptibility along both $[100]$ and $[001]$ directions, resulting in a total of nine scaleograms in one training sample. }
    \label{fig:cwt}
\end{figure}

(i) \textit{Magnetization} per single rare-earth ion, $\mu_a$, along direction $a$ as a function of external magnetic field $B_a$ for fixed temperature $T$. We choose a magnetic field range between $B_{\text{min}} = 0$~T and $B_{\text{max}} = 10$~T, and three temperatures $T_j = 1, 17, 300$~K to represent the behavior in typical field and temperature ranges which are easily accessible experimentally. We use $N^B_{\text{steps}} = 64$ equally spaced magnetic field points
\begin{align}
B_a = B_{\text{min}} + \frac{B_{\text{max}} - B_{\text{min}}}{N^B_{\text{steps}} - 1} \, b_a
\label{eq:3.1}
\end{align}
with $b_a \in \{0, \ldots, N^B_{\text{steps}}-1 \}$. Depending on the CF point symmetry group $\mathcal{G}$, we choose different high-symmetry directions---only one high-symmetry direction $a = [001] \equiv z$ (two high-symmetry directions, $a = \{[100], [001]\}$) for cubic (tetragonal and hexagonal) symmetry. Combined with the three temperature values $T_j$, this corresponds to three (six) sets of magnetization data: $\mu_a(B_a, T_j)$. Other choices of directions are of course possible, but we wanted to keep the size of the input data set as small as possible to keep the experimental work necessary to obtain it at a minimum.

(ii) \textit{Magnetic susceptibility} $\chi_a(T)$ along direction $a$ as a function of temperature $T$. We choose a temperature range between $T_{\text{min}} = 1$~K and $T_{\text{max}} = 300$~K using $N^T_{\text{steps}} = 64$ equally spaced temperature points
\begin{align}
T &= T_{\text{min}} + \frac{T_{\text{max}} - T_{\text{min}}}{N^T_\text{steps} - 1} \, t \,.
\label{eq:3.2}
\end{align}
We use the same directions $a$ for the susceptibility and the magnetization, corresponding to one (or two) sets of susceptibility data.

(iii) \textit{Magnetic specific heat} $c_M(T)$ as a function of temperature $T$. We use the same temperature range and step size as for the susceptibility, see \equref{eq:3.2}.

\begin{figure*}[tb]
  \centering
    \includegraphics[width=\linewidth]{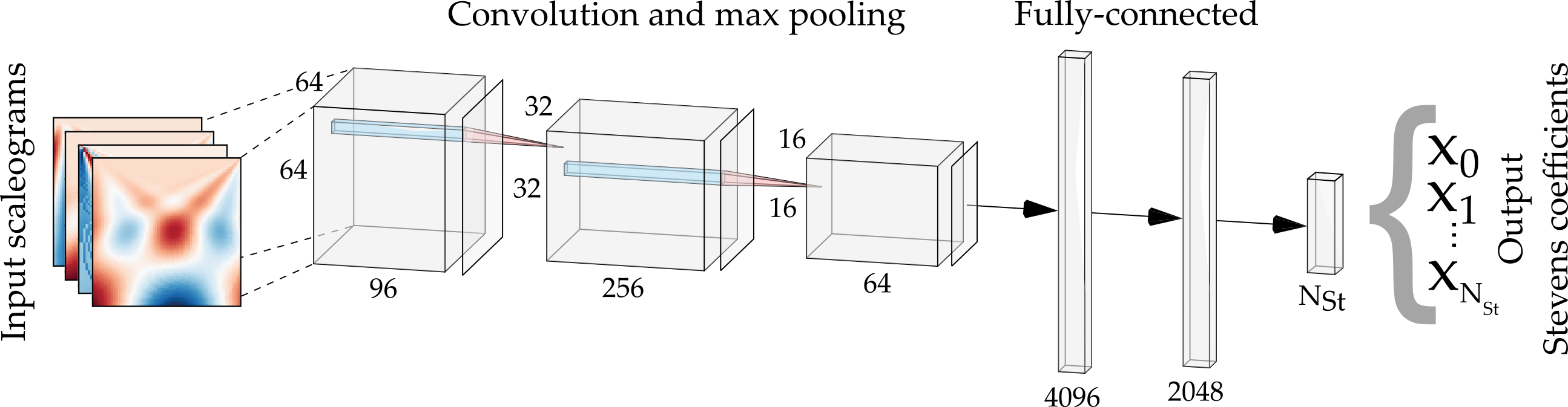}
    \caption{Schematic architecture of the 2D CNN that is used in this work. The multi-channel input image comprises five (nine) stacked CWT scaleograms of thermodynamic observables. The image first passes through three convolution and max-pooling layers. This allows the network to extract features from the images. The two following fully connected layers then predict the $N_{P} + 1$ Stevens coefficients, $\{x_0, x_1, ..., x_{N_{P}}\}$ in the output nodes. Here, $N_P = N_{\text{St}} - 1$ ($N_P = N_{\text{St}}$) for the cubic (hexagonal, tetragonal) case.  }
    \label{fig:architecture}
\end{figure*}
One training sample therefore consists of five (nine) different sets of thermodynamic data. A complete training sample for $J=4$, $\mathcal{G} = \text{m}\bar{3}\text{m}$ and $x_0 = 20$~K, $x_1 = 0.5$ is shown in Fig.~\ref{fig:training_data_set-J_4-Oh}(b, c).
To obtain the training data set, we draw the Stevens parameters randomly from a uniform
distribution and, for each of these sampled values, compute the aforementioned observables. To generate sufficient training data for the network, the process takes 2-3 hours. The wavelet transform described in the next subsection is included in this time frame. Note that within our convention there exist $N_{\text{St}}$ CF parameters for the cubic and $N_{\text{St}}+1$ CF parameters for the hexagonal and tetragonal cases. While $x_0$ can take either sign in the cubic case, it is strictly positive for hexagonal and tetragonal systems. In the latter cases, only the sign of the last Stevens parameter $\text{sign}(x_{N_{\text{St}}})$ enters the Hamiltonian.

\subsubsection{Continuous wavelet transform}
After comparing different network architectures (see more details below in Sec.~\ref{sec:convolutional_neural_network}), we have found it to be advantageous to first perform a continuous wavelet transformation (CWT) of the ``raw'' thermodynamic data before feeding it into a two-dimensional (2D) CNN. The reason is that CNNs are well suited to model data with an image-like structure like the wavelet scaleograms that are produced by the CWT. Similar to a Fourier transform, a CWT is used to perform a harmonic analysis and decompose a signal into its fundamental frequencies. The advantage of a CWT is that it produces a sparse representation of the data by providing localization in both frequency and ``time'' domain, with the main features of the data appearing in only a (small) subset of all CWT coefficients. This property is key for applications in data compression and denoising~\cite{mallatWaveletTourSignal2008}. We find that it also enables superior performance of a 2D CNN compared to placing a ``raw'' data vector of linear size $5 \times 64 = 320$ (or $9 \times 64 = 576$) on the network input nodes. Here, $5 (9)$ corresponds to the number of thermodynamic observables and $64$ to number $N^B_{\text{steps}}=N^T_{\text{steps}}$ of values of $B_a$ and $T$, respectively.

A CWT of a discrete and equally spaced 1D data set of size $N_f$,
\begin{equation*}
    \{f_0, f_1, \ldots f_{N_f-1}\} = \{f(t_\text{min}), f(t_{\text{min}} + \Delta), \ldots, f(t_{\text{max}})\}\,,
\end{equation*}
corresponds to performing the following transformation
\begin{equation}
    \tilde{f}(t, s) = \frac{1}{\sqrt{s}} \sum_{i=0}^{N_f-1} f_i \, \psi \Bigl( \frac{( i - t)\Delta}{s}\Bigr)\,.
\label{eq:3.4}
\end{equation}
Here, $\psi(t)$ is the so-called mother wavelet function, which is translated by parameter $t$ and scaled by the scale parameter $s$. The scale $s$ can be regarded as a period or inverse frequency. We choose a mother wavelet function of the real Morlet form~\cite{leePyWaveletsPywtPyWavelets2019}
\begin{equation}
    \psi(t) = e^{-\frac{t^2}{2}} \cos(5t). \label{MamaMorlet}
\end{equation}
We use a linearly spaced set of scales, $1 \leq s \leq 64$, which emphasizes the low frequency behavior of the data in comparison to using a geometric spacing. We have explicitly checked for a few cases that the CNN performs equally well if we use geometrically spaced scales. By calculating the convolution of this family of translated and scaled wavelets with our original data, we perform a frequency analysis that provides additional insight into the changes over the ``time'' domain, which corresponds to temperature or magnetic field in our data.

We apply a CWT to each of five (or nine) 1D observable data sets, $\{\mu_a(T,b_i), \chi_a(t_i), c_M(t_i)\}$, to produce a total of five (or nine) 2D wavelet scaleograms of size $64 \times 64$. In Fig.~\ref{fig:cwt}, we show four of the five scaleograms corresponding to the ``raw'' data in Fig.~\ref{fig:training_data_set-J_4-Oh}(b,c), which depict the associated (real) CWT coefficients $\tilde{f}$ defined in \equref{eq:3.4}. It clearly distinguishes regions with small and large wavelet coefficients, which is a characteristic of the underlying ``raw'' data set. The peak position and some broad characteristics of the original data, for example, whether a function approaches zero or a finite value at the boundary (minimal and maximal $T$ and $B$ values) can also be recognized in the scaleograms. Specifically, the peak position corresponds to a region with large wavelet coefficients at small scales $s$ (i.e. large frequencies), because the underlying function varies most rapidly close to the peak (see $\tilde{c}_M$ in Fig.~\ref{fig:cwt}, for example). Nonzero data values at boundaries (minimal and maximal $T$ or $B$ values) lead to pronounced peaks in the scaleogram at small $s$. The origin of this phenomenon is that zeros are padded to the dataset at both edges, which results in discontinuities at the boundary that show up as scaleogram peaks at small $s$. In Fig.~\ref{fig:cwt}, we can thus clearly distinguish observables that peak at low temperatures $\chi_z \propto 1/T$ (left boundary) from those that peak at the right boundary $\mu_{z,\text{T=17 K}}$. All five (nine) scaleograms are stacked into a multi-channel image and then placed on the input nodes of a 2D CNN, whose architecture is described in the next subsection.


\subsection{Convolutional neural network}
\label{sec:convolutional_neural_network}
We employ a CNN architecture that is based off the LeNet-5 architecture~\cite{Lecun98gradient-basedlearning}, which we scale up to be appropriate for the form of our training dataset. Initial experimentation with alternative architectures, such as simple feed-forward networks and 1D CNNs, yielded significantly worse results. To make a fair comparison, we created architectures that had approximately the same number of parameters as the 2D CNN and were trained on the same data. Applying to the cubic case with two Stevens parameters, $x_0$ and $x_1$, we find the 1D CNN a factor of 2 worse for $x_0$ and a factor of 7 worse for $x_1$ than the 2D CNN. The feed-forward deep neural network performed a factor of 1.25 worse for $x_0$ and a factor of $9$ worse for $x_1$ than the 2D CNN. It is expected that the performance difference is enhanced in the lower symmetry cases with more Stevens parameters to predict, which is why we chose to use the 2D CNN. As illustrated in Fig.~\ref{fig:architecture}, the input of the network is a five (nine) channel image containing the five (nine) scaleograms created using the CWT. We center the input data by subtracting the mean of each channel so that the distribution of the input ``pixel'' values has zero mean. We similarly normalize the target data $\{x_i\}$, as the coefficients $x_0$ and $x_{i}$ with $i \neq 0$ have significant size differences and different dimensions. Typically, $|x_0| \in [0.5, 50]$ (in units of K), while the dimensionless $x_i \in [-1,1]$ for $i >0$.

The centered input CWT scaleograms are fed into three sets of convolution and max-pooling layers. Each convolution layer has two identical 2D convolution layers with a kernel size of $3 \times 3$ and a stride of $1 \times 1$. This increases the number of channels and allows extracting data features. The max-pooling layers have a pool-size and stride of $2 \times 2$,
which essentially corresponds to a down-sampling of the image by a factor of two. Each convolution layer uses the ReLU activation function~\cite{DBLP:journals/corr/abs-1803-08375} and has batch normalization.

The final max-pooling layer is flattened and fed into two fully-connected layers, each with the ReLU activation function and $30\%$ dropout. These layers feed into a fully-connected output layer whose width corresponds to the number of independent Stevens parameters $N_{P} + 1$. Here, $N_P = N_{\text{St}}-1$ ($N_P = N_{\text{St}}$) for the cubic (hexagonal, tetragonal) case due to the procedure of splitting off the sign of $x_0$ into an additional parameter. The output layer uses a linear activation function. The $N_P+1$ output values are the prediction of the CNN for the Stevens coefficients $\{x_0, \ldots, x_{N_{\text{P}}}\}$,

The total number of trainable parameters in the network is $13,702,818 \approx 1.3 \times 10^7$. This is substantially larger than the number of trainable parameters in the original LeNet-5 architecture~\cite{Lecun98gradient-basedlearning}. This is due to the high dimensionality of the fully-connected layers in our version of the network. A larger number of parameters means that the network will be able to form more complex relationships between the features and the targets. However, we run the risk of over-parameterizing the network, resulting in a model that overfits---that is, it performs very well on the training data but poorly on unseen data. By applying normalization and dropout throughout the network we mitigate this issue. We build the network using Keras~\cite{chollet2015keras} and train it on Nvidia Volta V100S graphic processing unit (GPU). We use the Adam optimizer~\cite{DBLP:journals/corr/KingmaB14} with the recommended parameters to minimize the mean squared error (MSE) loss function,
\begin{equation}
    C(\hat{\mathbf{x}}, \mathbf{x}) = \frac{1}{(N_P+1) N_{\text{batch}}} \sum_{j=1}^{N_{\text{batch}}} \sum_{i=0}^{N_{P}} \bigl[\hat{x}_{i}(j) - x_{i}(j) \bigr]^2 \,, \label{LossFunction}
\end{equation}
where $\hat{x}_{i}$ is the network's prediction for $x_i$. We choose the stochastic gradient descent optimization algorithm \emph{Adam} to avoid as much as possible the trapping in local minima of the cost function. As shown below, the behavior of the quality of the CNN predictions (as described by MSE) across different input parameters can be largely understood on physical grounds such as arising from energy level crossings, from the smallness of certain CF parameters or from the ratio of the bandwidth to the maximal temperature scale. This indicates that the CNN is not trapped in local minima. In general, the inverse problem that the CNN addresses may be ill-defined and allow for multiple solutions. This issue can be (partially) addressed in practice by providing more data to the CNN such as enlarging the field and temperature range and/or by including magnetization and susceptibility data along different field directions.

Let us finally describe the resource cost of training the network. Using $10^5$ training examples and $1.5 \times 10^4$ validation and $1.5 \times 10^4$ testing examples with a batch size of $N_{\text{batch}} = 64$, the network converged after around $100$ epochs. With the available GPU (Nvidia Volta V100S, 32 GB), training the network took around 70 seconds per epoch. Fully training the network thus takes around 1-2 hours.


\section{CNN results} 
\label{sec:results}
In this section, we present results and measure the performance of CNNs predicting Stevens coefficients for three different point groups. We choose groups in cubic, hexagonal and tetragonal crystal classes that are of experimental relevance: $\text{m}\bar{3}\text{m}$, $\bar{6}\text{m}2$ and $4\text{mm}$. These groups allow for $N_{\text{St}} = 2, 4, 5$ independent Stevens parameters, respectively. The complexity of the task to find Stevens parameters from thermodynamic data increases when lowering the symmetry. We consider both integer and half-integer values of the total angular momentum quantum number $J$, and find that our method works equally well in both cases. For concreteness, we investigate $J=4$ and $J = 15/2$, which correspond to the ground state values of the rare-earth ions Pr$^{3+}$ ($J=4$) and Er$^{3+}$ ($J=15/2$).

For a given point group $\mathcal{G}$ and value of $J$, we train a CNN using the training data described in Sec.~\ref{sec:training_data_generation}. The input thus corresponds to five (nine) thermodynamic observables for cubic (hexagonal and tetragonal) point groups, and the output of the network are the $N_{P} + 1$ Stevens parameters, $\{x_0, \ldots, x_{N_P}\}$. We use two performance measures: (i) the \textit{mean absolute error} (MAE) of the network's prediction of the Stevens coefficients
\begin{align}
\text{MAE}(i) &= \frac{1}{N_{\text{test}}} \sum_{j=1}^{N_{\text{test}}} |\hat{x}_i(j) - x_i(j)|\,.
\label{eq:4.1}
\end{align}
Here, $\hat{x}_i$ is the prediction of the network,  $x_i$ is the true Stevens coefficient that is used to generate the data placed on the input nodes, and $N_{\text{test}}$ is the size of the testing dataset. The MAE is related to the loss function (\ref{LossFunction}) used to train the network.
(ii) The \textit{mean squared error} (MSE) of two thermodynamic data sets $\{\mu_a, \chi_a, c_M\}$ generated by $\hat{x}_i$ and $x_i$, respectively:
\begin{align}
    \text{MSE}(\{x_i\}) &= \frac{1}{M} \sum_{\nu = 1}^M \bigl[ \bar{\mathcal{O}}_\nu(\{ \hat{x}_i\}) - \bar{\mathcal{O}}_\nu(\{ x_i\})  \bigr]^2 \,.
\label{eq:4.2}
\end{align}
 Here, $M = 5 \times 64 = 320$ ($9 \times 64 = 576$) for cubic (hexagonal, tetragonal) point groups is the length of the thermodynamic dataset and $\bar{\mathcal{O}}_\nu$ runs over the five (nine) experimental observables $\{\mu_{a, T_\alpha}(b_i), \chi_a(t_i), c_M(t_i)\}$ as a function of temperature $t_i$ and magnetic field $b_i$ (see Sec.~\ref{sec:training_data_generation}) that are obtained for a given choice of Stevens parameters. To account for the differences in size and units between observables, we first normalize each dataset by their mean and perform Eq.~\ref{eq:4.2} on the resulting dimensionless quantities.

The MSE measures the performance of the CNN in reproducing the desired (input) thermodynamic data set that was generated using $\{x_i\}$. We include this metric as the sensitivity of the error in the observables (MSE) with respect to the error in the Stevens parameters (MAE) depends on the values of the $\{x_i\}$, and the MSE thus contains additional information about the networks performance.  Unless otherwise noted, both MAE and MSE are evaluated on a testing data of size $N_{\text{test}} = 4\times 10^3$ that was not shown to the network during training.

In the following, we separately discuss the performance of the CNNs for the cubic, hexagonal and tetragonal point groups.

\subsection{Cubic point group symmetry} 
\label{sub:cubic_point_group_symmetry}
\begin{figure}[tb]
    \centering
    \includegraphics[width=1\linewidth]{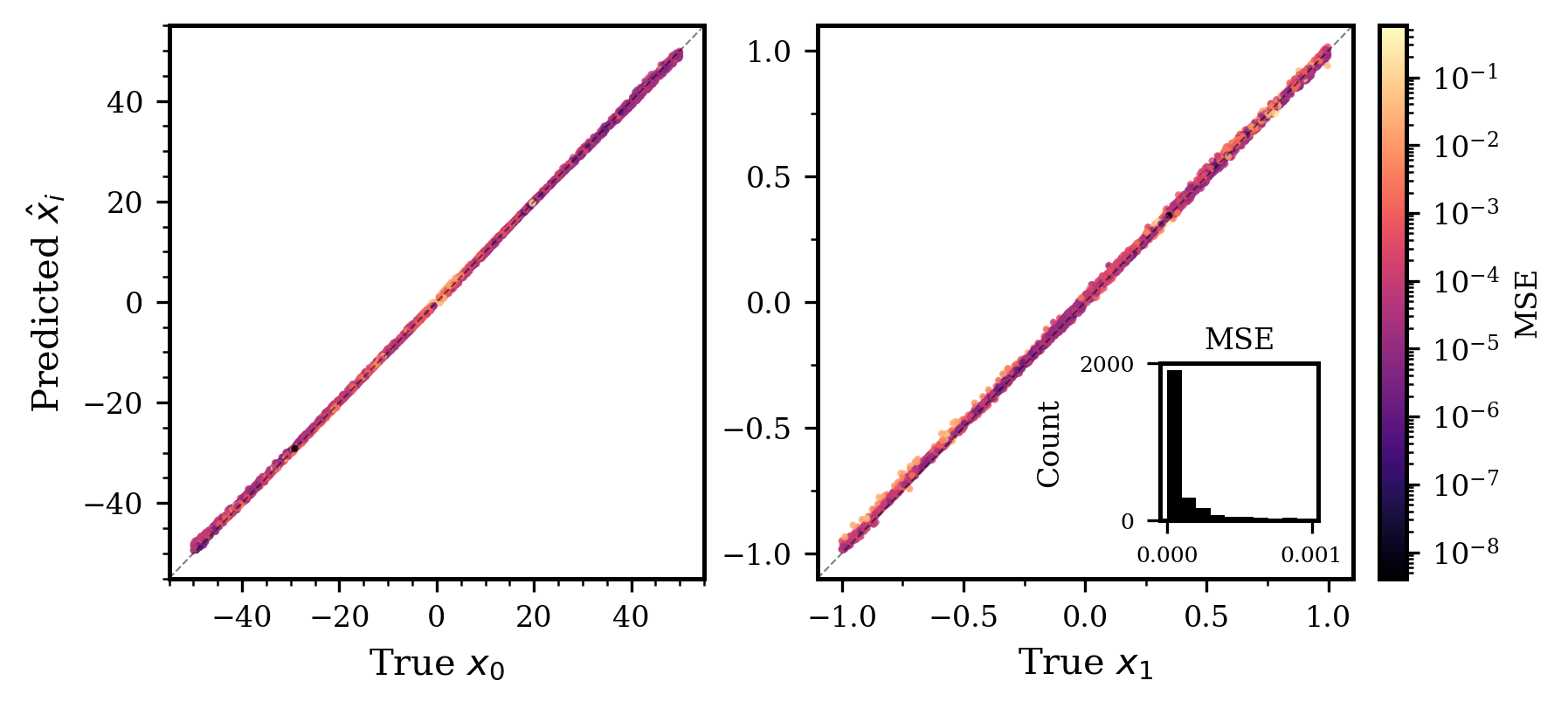}
    \caption{True Stevens coefficients $x_i$ versus predicted coefficients $\hat{x}_i$ for cubic point group $\text{m}\bar{3}\text{m}$ and $J=4$. The network was trained with the five thermodynamic data sets $\{\mu_{z,T_\alpha}(b_i), \chi_z(t_i), c_M(t_i)\}$ for $T_\alpha = \{1, 17, 300\}$~K. We choose the high-symmetry direction $z = [001]$. The Stevens coefficients are predicted with MAEs given by $\text{MAE}(0) = 0.321$~K and $\text{MAE}(1) = 0.012$. The inset and color coding shows the MSE of each data point, which lies at $\braket{\text{MSE}} = 0.053$ on average. }
    \label{fig:Oh4resultscomb}
\end{figure}

We consider the case of cubic point group $\mathcal{G} = \text{m}\bar{3} \text{m}$ and $J = 4$, which is applicable to cubic Pr rare-earth compounds. The energy level diagram for this case is shown in Fig.~\ref{fig:training_data_set-J_4-Oh}(a) and exhibits singlet, doublet and triplet ground states, depending on the sign of $x_0$ and the value of $x_1$. As shown in Fig.~\ref{fig:Oh4resultscomb}, the CNN accurately predicts the two independent Stevens coefficients $\{x_0, x_1\}$ with error values of $\text{MAE}(0) = 0.321~\text{K}$ and $\text{MAE}(1) = 0.012$. Note that we choose the energy range of $0.5~\text{K} \leq |x_0| \leq 50$~K. The color code and the inset show the MSE, which lies at $\braket{\text{MSE}} = 0.053$ on average. The results in Fig.~\ref{fig:Oh4resultscomb} show the predictions of two networks: one is trained with strictly positive $x_0 \in [0.5, 50]$, and a second one is trained with strictly negative $x_0 \in [-50, -0.5]$. When applied to a given testing example, which has a definite but unknown sign of $x_0$, the performance of the network that was trained on data with the same sign of $x_0$ as the testing example is typically much better and can be easily identified. Here, we show results for testing examples that have a known sign for simplicity, i.e., the positive (negative) network is tested on samples with positive (negative) $x_0$.

\begin{figure}[tb]
    \centering
    \includegraphics[width=0.9\linewidth]{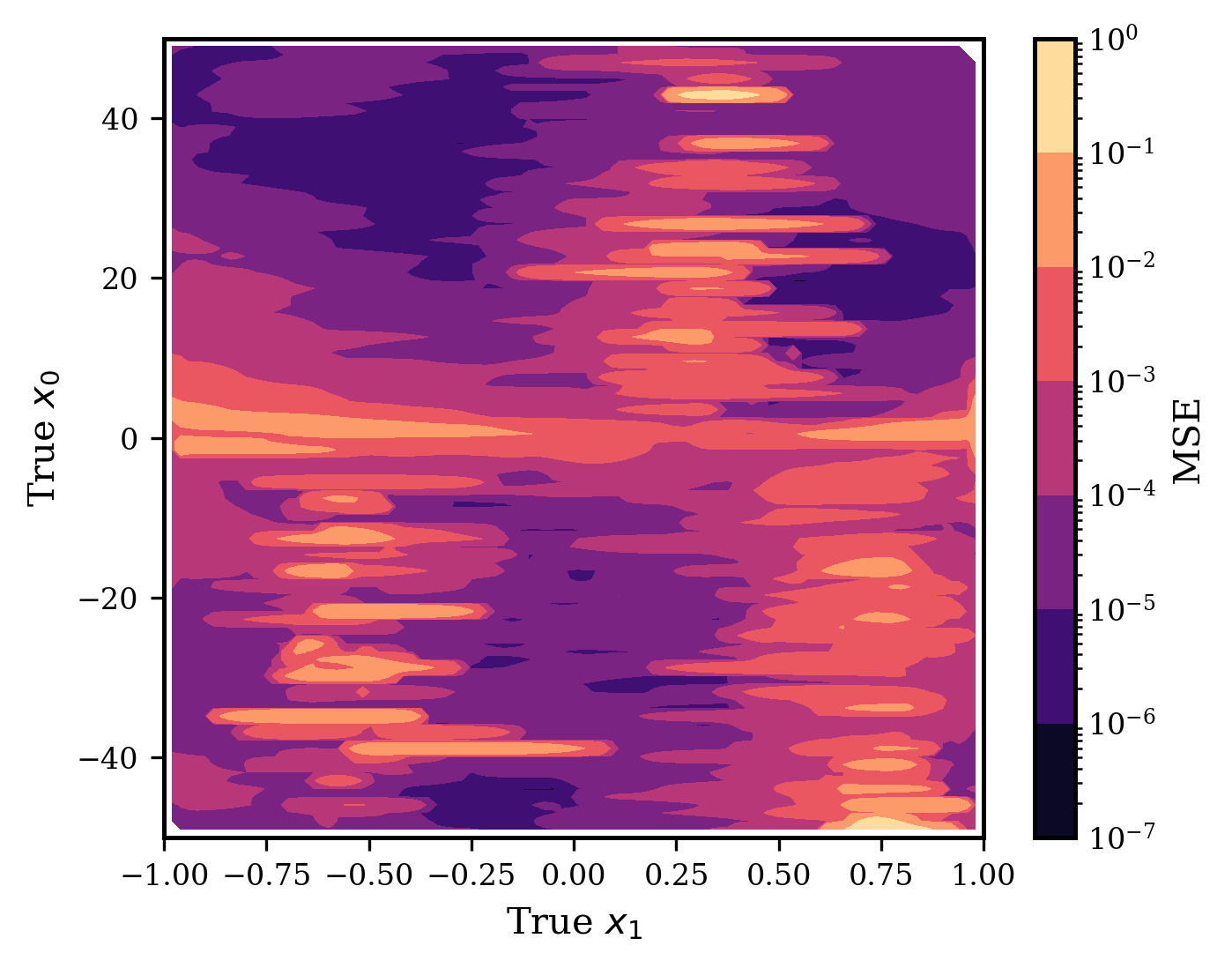}
    \caption{Heat map showing the MSE as a function of the true Stevens coefficients $x_0$ and $x_1$ for $\mathcal{G} = \text{m}\bar{3} \text{m}$ and $J=4$. The heat map shows a total number of $N_{\text{testing}} = 4 \times 10^3$ data points. Regions with larger MSE occur when $x_0$ becomes comparable to the temperature probed ($x_0 \lesssim T_{\text{min}} = 1$~K) and when there is an energy level crossing involving the ground state [see Fig.~\ref{fig:training_data_set-J_4-Oh}(a)]. While it becomes impossible to predict the coefficients if $x_0 \lesssim T_{\text{min}}$ as the spectrum collapses, the increased MSE at the position of level crossings rather indicate an enhanced sensitivity of the observables with respect to small errors in the $\{\hat{x}_i\}$, which are still accurately predicted by the network (see Fig.~\ref{fig:Oh4resultscomb}). }
    \label{fig:oh4}
\end{figure}

In Fig.~\ref{fig:oh4}, we visualize the distribution of MSE as a function of the two Stevens parameters $x_0$ and $x_1$. We clearly observe that the MSE is larger in regions where $x_0$ is small. This occurs as the energy level spectrum collapses in this limit, with all levels being smaller than (or comparable to) the minimal thermal energy $\approx k_B T_{\text{min}}$ at $T_{\text{min}} = 1$~K. In this regime, thermodynamic data cannot  resolve the order of the levels. We also find an increased MSE along the lines $x_1 \approx 0.35$ for positive $x_0 > 0$ and $x_1 = -0.6$ and $x_1 = 0.75$ for negative $x_0 < 0$. This follows from the fact that the ground state energy exhibits level crossings in these parameter regions, as shown in Fig.~\ref{fig:training_data_set-J_4-Oh}(a). This makes the thermodynamic observables quite sensitive to small errors in the Stevens parameters as the nature of the ground state changes between singlet, doublet and triplet states. As a result, the MSE is enhanced even though the MAE is still small and the Stevens coefficients are predicted with high accuracy.

\subsection{Hexagonal point group symmetry} 
\label{sub:hexagonal_point_group_symmetry}
We also consider the case of hexagonal point group $\mathcal{G} = \bar{6}\text{m}2$ and $J = 15/2$, which is applicable to hexagonal Er rare-earth compounds. Being a half-integer value of $J$, the energy level exhibits Kramers degeneracy in the absence of a magnetic field. The number of independent Stevens coefficients for this point group is $N_{\text{St}} = 4$ (see Eq.~\eqref{eq:2.11}). We split off the sign of $x_0$ into a separate parameter $x_4$ such that $\text{sign}(x_4) = \text{sign}(x_0)$. This allows us to consider the parameter $x_0 \equiv |x_0|$ to be strictly positive. The training data sets thus contains the five coefficients, $\{x_0, \ldots, x_4\}$, with strictly positive $x_0 > 0$ and only the sign of $x_4$ entering the Hamiltonian. We consider the scale parameter to be in the region $0.5~\text{K} \leq x_0 \leq 50~\text{K}$ and $-1 \leq x_i \leq 1$ for $i \geq 1$.
\begin{figure*}[tb]
    \centering
    \includegraphics[width=\linewidth]{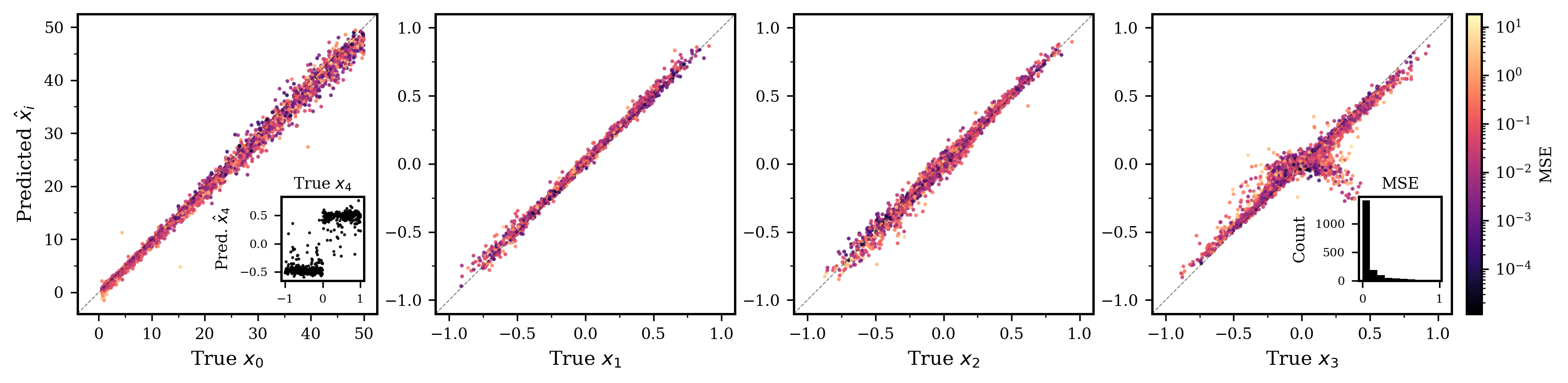}
    \caption{True Stevens coefficients $x_i$ versus predicted coefficients $\hat{x}_i$ for hexagonal $\bar{6}\text{m}2$ point group and $J=15/2$ for $N_{\text{testing}} = 4000$ data sets. The color denotes the MSE and the inset histogram in the right most panel shows the distribution of the MSE. This network was trained using nine 1D observable data sets, $\{\mu_a(T,b_i), \chi_a(t_i), c_M(t_i)\}$. The MAEs of the Stevens coefficients are $\{\text{MAE}(0),\ldots, \text{MAE}(3)\} = \{1.325~\text{K}, 0.024, 0.031, 0.073\}$. The sign of $x_4$ was correctly predicted in 96\% of the cases (see inset in left most panel). The average MSE is $\braket{\text{MSE}} = 0.280$. }
    \label{fig:d3h7.5}
\end{figure*}

As described in Sec.~\ref{sec:training_data_generation}, the training data contains in addition to the specific heat $c_M$, the magnetization $\mu_a$ and susceptibility $\chi_a$ along both $a=[100]$ and $a=[001]$ directions. This provides information about the anisotropy between the $ab$ plane and the $c$ axis in the system, and is necessary for the CNN to be able to predict the parameter $x_3$. A training data set thus consists of the nine observables $\{\mu_{a,T_\alpha}(B_a), \chi_{a}(T), c_M(T)\}$. Here, we set $T_\alpha = 1, 17, 300$~K and consider the magnetic field range $0 \leq B_a \leq 10$~T and temperature range $1~\text{K} \leq T \leq 300$~K as described in Sec.~\ref{sec:training_data_generation}.

As shown in Fig.~\ref{fig:d3h7.5}, the CNN can accurately predict the Stevens parameters with $\{\text{MAE}(0), \ldots, \text{MAE}(3)\} = \{ 1.325~\text{K}, 0.024, 0.031, 0.073\}$. The sign of the fifth parameter $x_4$ was correctly found in 96\% of the cases (see inset in the left panel in Fig.~\ref{fig:d3h7.5}). The color corresponds to the MSE of each testing data set. A histogram of the MSE values is included in the right most panel. The average MSE over all testing data sets is $\braket{\text{MSE}} = 0.280$.
In general, the MAE increases slightly with larger values of $x_0$, which can be understood from the fact that the thermal energy is not sufficient to probe higher lying levels. This could likely be improved by enlarging the temperature range by increasing $T_{\text{max}}$. The CNN performs worst for the $x_3$ coefficient, in particular when this parameter is small. This parameter contains information about the anisotropy between $ab$ and $c$ axis directions as well as between directions within the $ab$-plane. The CNN predictions of $x_3$ could thus likely be improved by providing additional magnetization data along a second, inequivalent direction in the $ab$ plane.
Finally, we note that quantitatively similar results were obtained for the integer case of $J=4$, showing that the method works equally well for integer and half-integer values of $J$.


\subsection{Tetragonal point group symmetry} 
\label{sub:tetragonal_point_group_symmetry}
We study the performance of the CNN for the tetragonal point group $\mathcal{G} = 4\text{mm}$ and half-integer $J = 15/2$, which corresponds to tetragonal Er rare-earth systems. In this case, the CF allows for $N_{\text{St}} = 5$ independent Stevens parameters. Since we split off the sign of $x_0$, the CNN actually predicts six parameters $\{x_0, \ldots, x_5\}$, where $0.5 \leq x_0 \leq 50$ (in units of K), $-1 \leq x_i \leq 1$ for $i \geq 1$ and the training data depend only on the sign of $x_5$. A training data set contains nine scaleograms obtained from specific heat $c_M$, magnetization $\mu_a$ and susceptibility $\chi_a$ along $a = [100], [001]$ directions: $\{\mu_{a, T_\alpha}(B_a), \chi_a(T), c_M(T)\}$. The temperature and field ranges are $1~\text{K} \leq T \leq 300~\text{K}$ and $0 \leq B_a \leq 10$~T. Providing information about the anisotropy between the $ab$ plane and the $c$ axis is necessary for the CNN to be able to learn the dependence on the parameters $x_2$ and $x_4$.
\begin{figure*}[tb]
    \centering
    \includegraphics[width=\linewidth]{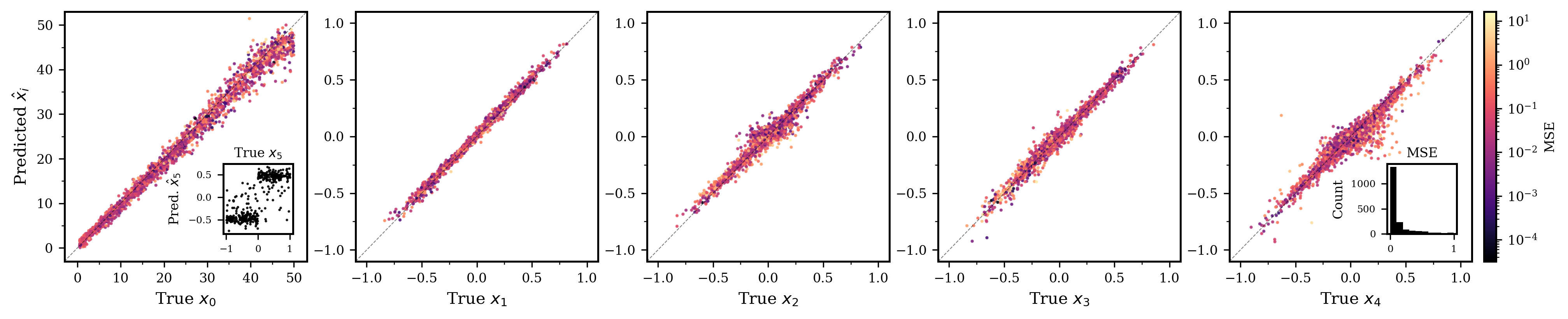}
    \caption{True Stevens coefficients $x_i$ versus predicted coefficients $\hat{x}_i$ for tetragonal $4\text{mm}$ point group and $J=15/2$ for $N_{\text{testing}} = 4000$ data sets. The color denotes the MSE and the inset histogram in the right most panel shows the distribution of the MSE. This network was trained using nine 1D observable data sets, $\{\mu_a(T,b_i), \chi_a(t_i), c_M(t_i)\}$. The MAEs of the Stevens coefficients are $\{\text{MAE}(0),\ldots, \text{MAE}(4)\} = \{1.380~\text{K}, 0.022, 0.038, 0.031, 0.059\}$. The sign of $x_5$ was correctly predicted correctly in 93\% of the cases (see inset in left most panel). The average MSE is $\braket{\text{MSE}} = 0.248$. }
    \label{fig:c4v7.results5}
\end{figure*}

As shown in Fig.~\ref{fig:c4v7.results5}, the CNN can accurately predict the Stevens parameters for the majority of the data points that it was tested on. The MAEs of the Stevens coefficients read $\{\text{MAE}(0),\ldots, \text{MAE}(4)\} = \{1.380~\text{K}, 0.022, 0.038, 0.031, 0.059\}$. The sign of $x_5$ was correctly predicted by the network in $93\%$ of the cases. The average MSE is given by $\braket{\text{MSE}} = 0.248$. The overall performance is comparable to the hexagonal case of $\bar{6}\text{m}2$, even though the tetragonal case exhibits one more Stevens parameter. Similar to the hexagonal case, the error is larger for larger values of $x_0$, which likely stems from the fact that the bandwidth of the spectrum becomes larger than the maximal thermal energy $k_B T_{\text{max}}$. This suggests increasing the temperature range in the training data. The MAE of different coefficients is comparable. The largest MAE occurs for the parameter $x_4$, which measures the anisotropy of the system, both between $ab$ and $c$ directions as well as within the $ab$ plane. It could likely be better predicted by adding magnetization data along another inequivalent direction in the $ab$ plane to the training sets. Finally, we note that we have applied the algorithm to the case of $J=4$ and obtained quantitatively similar results.



\section{Application to experimental data} 
\label{sec:application_to_experimental_data}
The ultimate application of the presented CNN algorithm is to extract CF parameters from real experimental data. We provide all required programs as open-source software~\cite{orth_berthusen_2020}. In this section, we demonstrate this by applying the algorithm to three published experimental data sets: one Cerium and two Praseodymium-based rare-earth intermetallics: (i) CeAgSb$_2$~\cite{Takeuchi-PRB-2003, MYERS199927} and (ii) PrAgSb$_2$~\cite{MYERS199927}, where the rare-earth ions Ce$^{3+}$ ($J=5/2$) and Pr$^{3+}$ ($J=4$) experience $\text{4mm}$ site symmetry, and (iii) PrMg$_2$Cu$_9$~\cite{PhysRevB.94.144434}, where Pr$^{3+}$ exhibits $\bar{6}\text{m}2$ site symmetry. Importantly, published data of magnetization, magnetic susceptibility and magnetic specific heat on the same single crystal are available for these systems~\cite{Takeuchi-PRB-2003, MYERS199927,PhysRevB.94.144434}. Since experimental circumstances and parameters are different for each material, it is required to train a custom CNN for each case.

When applying the CNN algorithm to experimental data, one must first select the set of thermodynamic observables that are given to the network. In general, it is best to include as much data as possible, for example, magnetization along different directions, which informs the network about the anisotropy in the material. In addition, one must choose the temperature and magnetic field ranges. These will in general be different for each observable to ensure that the assumptions of the modeling, in particular the single-ion approximation, are valid. Once the set of experimental input data is determined (together with $J$ and $\mathcal{G}$), one generates a customized training data set using the same observables and parameter ranges. Finally, to use the experimental data as input data for the CNN, one first needs to transform from the experimental units to the units used in the training data. This is described in detail in Appendix~\ref{app:sec:units_transformation}.

\begin{figure*}[tbh]
    \centering
    \includegraphics[width=\linewidth]{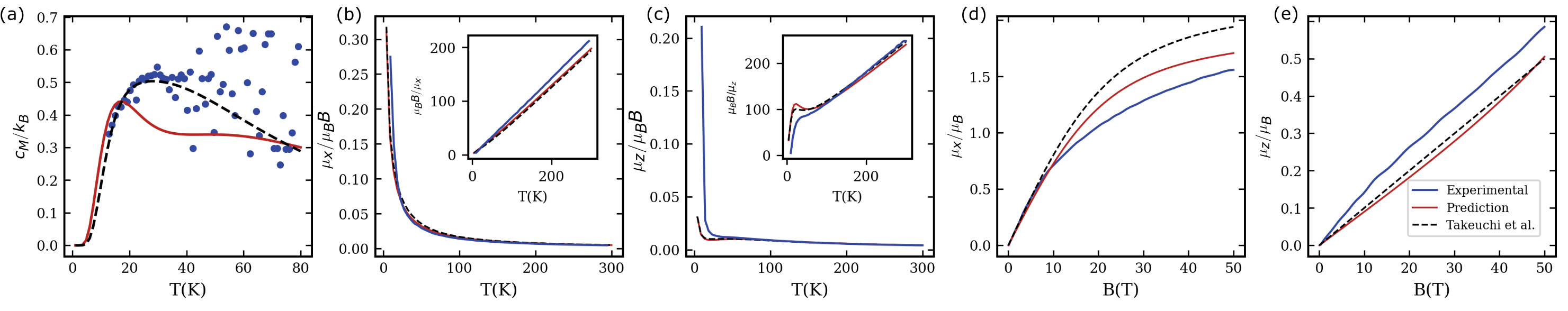}
    \caption{CNN application to CeAgSb$_2$. Comparison of experimental target data and CNN predictions for CeAgSb$_2$. Experimental target data is shown in blue. The theoretical results are obtained within the single-ion approximation using the CF parameters in Tab.~\ref{tab:0} obtained from CNN  (red) and from Takeuchi \emph{et al.}~\cite{Takeuchi-PRB-2003} (black dashed). Both fits are characterized by  $\text{MSE} = 0.17$. }
    \label{fig:exp_CeAgSb2}
\end{figure*}

\begin{figure}[tbh]
    \centering
    \includegraphics[width=\linewidth]{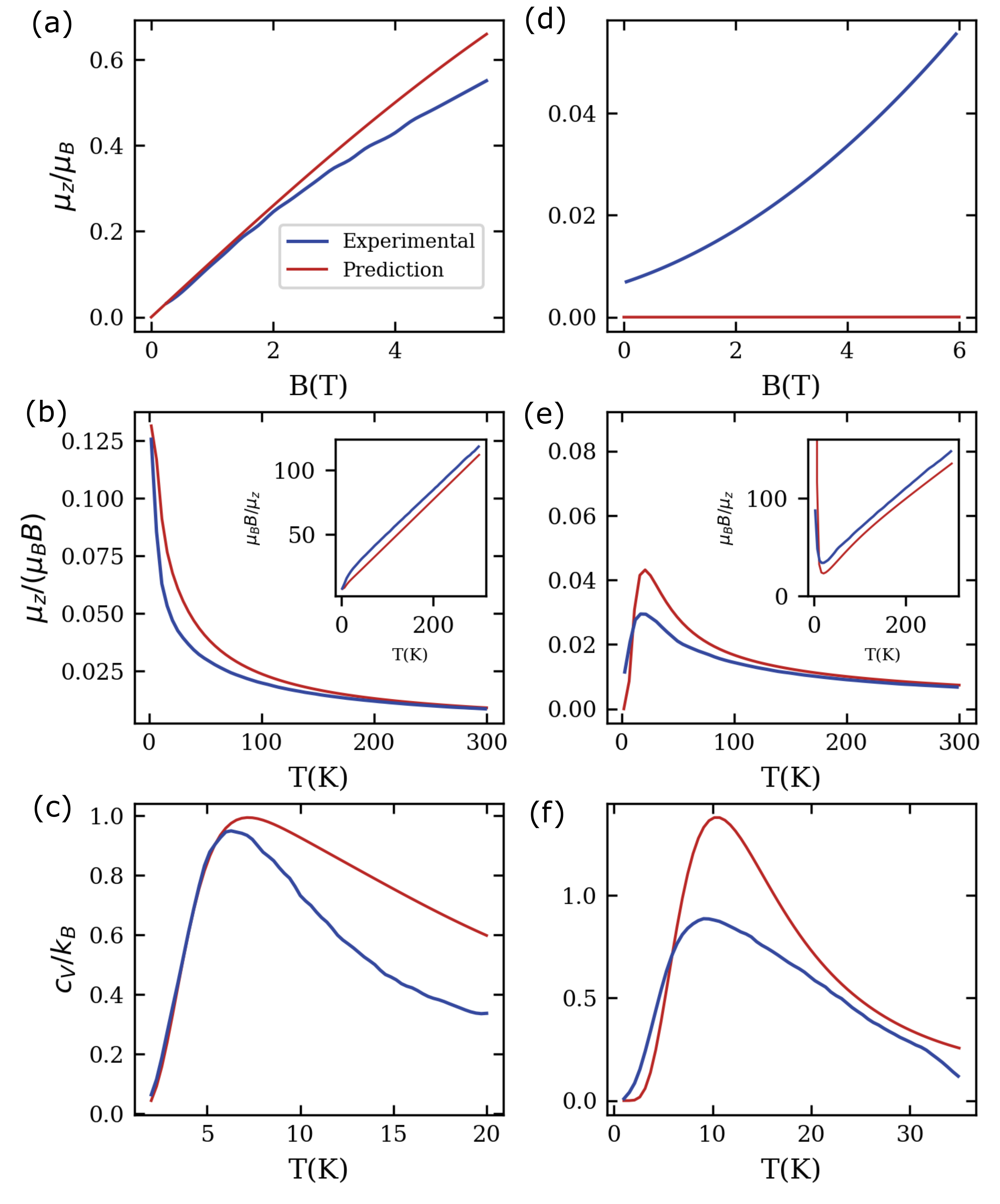}
    \caption{CNN application to PrAgSb$_2$ (a-c) and PrMg$_2$Cu$_9$ (d-f). Panels show experimental data (blue) and CNN predictions (red) for input data $\{\mu_{z,\text{T} = 2\,\text{K}}, \chi_z(T), c_M(T)\}$. Experimental data is from Refs.~\cite{MYERS199927,PhysRevB.94.144434}. Theory results are calculations (within the single-ion approximation ) using the CF parameter values $\{\hat{x}_i\}$ predicted by the CNN (see Tab.~\ref{tab:1} and~\ref{tab:2}). We find good agreement between experimental data and CNN predictions with a $\text{MSE} = 0.074$ (PrAgSb$_2$) and MSE = 0.085 (PrMg$_2$Cu$_9$). Note that due to the lack of experimental data containing information about the magnetic anisotropy, the parameters $x_2, x_4$ (for PrAgSb$_2$) and $x_3$ (for PrMg$_2$Cu$_9$) cannot be properly learned by the CNN. Also note the small $y$ axis scale in panel (d) due to a strong $ab$ easy-plane anisotropy.}
    \label{fig:exp_res_combined}
\end{figure}

\begin{table}[tb]
    \centering
    \begin{tabular}{|c|c||c|c|c|}
    \hline
     & $x_i$ & $(k,q)$ & $\mathcal{B}^q_k$& $B^q_{k, \text{Stevens}}$  \\
    \hline
    $x_0$ & $29$~K & -- & -- & -- \\
    \hline
    $x_1$ & $0.61$ & $(2,0)$ & $18.4$~K & $5.8$~K\\
    \hline
    $x_2$ & $0.80$ & $(4,4)$ & $2.9$~K & $2.6$~K \\
    \hline
    sign($x_3$) & $-1$ & $(4,0)$ & $2.1$~K & $0.22$~K \\
    \hline
    \end{tabular}
    \caption{Stevens parameters for CeAgSb$_2$ obtained from CNN using input experimental data from Ref.~\cite{Takeuchi-PRB-2003} of observables $\{\mu_{\alpha,T=20\,\text{K}}, \chi_\alpha(T), c_M(T)\}$ for directions $\alpha = \{x \equiv [100], z \equiv [001]\}$. The coefficients $x_i$, $\mathcal{B}^q_k$ and $B^q_{k, \text{Stevens}}$ are defined in Sec.~\ref{sec:crystal_field_thermodynamics}.
    Note that Takeuchi \emph{et al.} in Ref.~\cite{Takeuchi-PRB-2003} report the Stevens coefficient values $B^0_{2, \text{St.}} = 7.55$~K, $B^4_{4, \text{St.}} = -0.64$~K and $B^0_{4, \text{St.}} = -0.02$~K. Another set of values (closer to our findings) is reported by Jobiliong \emph{et al.} in Ref.~\cite{Jobiliong-PRB-2005}: $B^0_{2, \text{St.}} = 6.60$~K, $B^4_{4, \text{St.}} = 1.14$~K and $B^0_{4, \text{St.}} = -0.09$~K.}
    \label{tab:0}
\end{table}

\begin{table}[tb]
    \centering
    \begin{tabular}{|c|c||c|c|c|}
    \hline
     & $x_i$ & $(k,q)$ & $\mathcal{B}^q_k$& $B^q_{k, \text{Stevens}}$  \\
    \hline
    $x_0$ & $39$~K & -- & -- & -- \\
    \hline
    $x_1$ & $0.19$ & $(2,0)$ & $4.1$~K & $1.3$~K\\
    \hline
    $x_2$ & -- & $(4,4)$ & -- & -- \\
    \hline
    $x_3$ & $-0.32$ & $(4,0)$ & $-0.39$~K & $-0.041$~K \\
    \hline
    $x_4$& -- & $(6,4)$ & -- & -- \\
    \hline
    sign($x_5$) & $-1$ & $(6,0)$& $-0.038$~K & $-2.4 \times 10^{-3}$~K\\
    \hline
    \end{tabular}
    \caption{Stevens parameters for PrAgSb$_2$ obtained from CNN using $\{\mu_{z,T=2\,\text{K}}, \chi_z(T), c_M(T)\}$ from Ref.~\cite{MYERS199927} as input data. The coefficients $x_i$,  $\mathcal{B}^q_k$, and $B^q_{k, \text{Stevens}}$ are defined in Sec.~\ref{sec:crystal_field_thermodynamics}.
    Note that the CNN is unable to learn the coefficients $\{x_2, x_4\}$ since the experimental data set (and thus also the training data sets) only include susceptibility and magnetization along the $z$ direction and thus does not include sufficient information about the magnetic anisotropy. This is a limitation of the experimental data set, and not of our deep learning approach, as shown in Sec.~\ref{sec:results}, where data along a second axis is included and $\{x_2, x_4\}$ are predicted well. Finally, we note that Myers \emph{et al.} report $B^0_{2, \text{Stevens}} = 1.8 \pm 0.3$~K~\cite{MYERS199927}, close to what we find. }
    \label{tab:1}
\end{table}

When selecting a suitable temperature window, one must ensure to avoid the occurrence of many-body phenomena such as the development of Kondo screening (by choosing $T > T_K$), magnetic order ($T > T_M$), or significant magnetic exchange interaction effects ($T > T_{\text{RKKY}}$), which are currently neglected in the modeling that generates the training data. We thus choose to apply our algorithm to the Praseodymium members of the RAgSb$_2$ and RMg$_2$Cu$_9$ series, because they do not exhibit magnetic order down to $2$~K (even though magnetic exchange effects may become noticeable at $T \lesssim 5-10$~K already). On the other hand, they may exhibit some degree of $J$ mixing~\cite{kuzminChapterThreeTheory2007,Princep-PRB-2013}, which is currently neglected in the training data generation. As this is avoided in the Ce member of the series, because Ce$^{3+}$ only contains a single $f$ electron, we also investigate CeAgSb$_2$ within our approach.

When including specific heat data, it is also important to realize that the magnetic part of $c_M$ is typically experimentally approximated by subtracting off the specific heat of a corresponding nonmagnetic compound. A nonmagnetic analogue material can often be obtained by replacing the magnetic rare-earth ion by a nonmagnetic one such as La, Y or Lu. The subtraction procedure is only valid when all other (i.e., the phonon and electronic) contributions to the specific heat in the two materials are identical. In practice, this restricts the temperature regime that can be used for (magnetic) $c_M$ in the algorithm. Similarly, one may need to subtract an enhanced Pauli susceptibility contribution, which arise from conduction electrons, from the magnetic susceptibility data.

We emphasize that these caveats are related to the physical modeling of the forward problem of computing (or experimentally extracting) observables. Our CNN-based approach of solving the inverse problem, however, is more generally valid and can, in principle, also be used in conjunction with more advanced and realistic physical models (that might, e.g., be able to capture magnetism or phonons). Magnetic exchange interactions could be incorporated rather straightforwardly within a molecular mean-field approach~\cite{Takeuchi-CeRhIn5-2001, Jobiliong-PRB-2005}.

Based on these general considerations, we select Ce and Pr members of the RAgSb$_2$ series and PrMg$_2$Cu$_9$ as suitable experimental systems to apply and test our deep learning algorithm in practical situations. The material CeAgSb$_2$ develops magnetic order at $T=9.7$~K~\cite{Takeuchi-PRB-2003} and we thus restrict the temperature regime for which we generate training data to be between $10~\text{K} \leq T \leq 300~\text{K}$. In contrast, both Pr compounds that we investigate remain paramagnetic down to $T=2$~K and can be well described within the single-ion approximation over the complete temperature range from $T_{\text{min}} = 2$~K to $T_{\text{max}} = 300$~K.

For all three compounds, published data exists for magnetic specific heat $c_M$ as well as magnetization and susceptibility along the $[001]$ axis (and also along the $[100]$ axis in case of CeAgSb$_2$). We note that while there exists data for the Pr compounds in Refs.~\cite{MYERS199927, PhysRevB.94.144434} with magnetic fields applied in the $ab$ plane, these will not be included, because the exact in-plane direction was not experimentally determined. This implies that the Stevens coefficients that describe the anisotropy in the $ab$ plane cannot be determined for the Pr compounds.

\subsection{CeAgSb$_2$}
\label{subsec:CeAgSb2}
The thermodynamic properties of tetragonal material CeAgSb$_2$ were studied in detail in Ref.~\cite{MYERS199927, Takeuchi-PRB-2003,Jobiliong-PRB-2005}. This Ce-based Kondo lattice system orders ferromagnetically below $T_C = 9.6$~K with moments aligned parallel to the $c$ axis. Crystal fields were previously shown to play an important role in the material, leading to a peculiar magnetization behavior with moments ordering along the magnetically hard ($c$) axis, and  saturation moments that are larger for fields lying in the easy ($ab$) plane (versus the $c$ axis)~\cite{MYERS199927,Takeuchi-PRB-2003, Araki-PRB-2003, Hafner-PRB-2019}.

We use experimental results reported in Ref.~\cite{Takeuchi-PRB-2003} as input data for the CNN. In Fig.~\ref{fig:exp_CeAgSb2}, we show the complete experimental data set provided to the CNN (after unit conversion and wavelet transformation as described in Secs.~\ref{app:sec:units_transformation} and~\ref{sec:convolutional_neural_network_approach_to_finding_cf_parameters}). It contains the magnetic specific heat $c_M$ between $13$~K and $80$~K, and magnetic susceptibility (versus $T$) and magnetization (versus $B$ at $T=20$~K) with fields applied along the $x \equiv [100]$ and $z \equiv [001]$ axes. The figure compares the experimental data to the theoretical results obtained within the single-ion approximation that uses the values  of Stevens coefficients $\{\hat{x}_i\}$ predicted by the CNN (red) and reported by Takeuchi \emph{et al.} in Ref.~\cite{Takeuchi-PRB-2003} (black dashed).

Overall, we find that the CNN predictions match the experimental data very well with a $\text{MSE} = 0.17$. The results obtained with values from Ref.~\cite{Takeuchi-PRB-2003} are characterized by the same $\text{MSE}$. While the values from Ref.~\cite{Takeuchi-PRB-2003} provide a slightly better fit to (the noisy data of) $c_M$, the CNN predictions lead to a slightly improved fit of $\mu_x(B)$ (see Fig.~\ref{fig:exp_CeAgSb2}(d)). Ultimately, a better fit may require including effects of spin exchange, as done in Ref.~\cite{Jobiliong-PRB-2005} via a molecular field approach. Such an approach could be straightforwardly included in our algorithm, which is left for future work.
Table~\ref{tab:0} contains the values of the Stevens coefficients from the CNN, and compares them to two sets reported in the literature~\cite{Takeuchi-PRB-2003,Jobiliong-PRB-2005}. The values obtained from our CNN approach are closer to those reported in Ref.~\cite{Jobiliong-PRB-2005}.
In all three sets, the coefficient $B^0_{2,\text{Stevens}} \equiv B^0_{2,\text{St.}}$ is the dominant CF coefficient, followed by $B^4_{4,\text{St.}}$. We note that $B^0_{2,\text{Stevens}}$ is proportional to the difference in Curie-Weiss temperatures along different axes~\cite{Wang-Phys_Lett_A-1971}, which provides another useful validation check of the CNN results.

\subsection{PrAgSb$_2$ and PrMg$_2$Cu$_9$}
\label{subsec:PrAgSb2_PrMg2Cu9}
We apply our CNN algorithm to two Pr based materials: tetragonal PrAgSb$_2$ and hexagonal PrMg$_2$Cu$_9$. Both materials remain paramagnetic down to $T = 2$~K, and are modelled within the single-ion approximation over the full temperature range from $T = 2$ to $T = 300$~K.

In Fig.~\ref{fig:exp_res_combined}, we show results of our CNN algorithm together with the experimental data for PrAgSb$_2$ (a-c) and PrMg$_2$Cu$_9$ (d-f). The experimental data is taken from Refs.~\cite{MYERS199927} (PrAgSb$_2$) and \cite{PhysRevB.94.144434} (PrMg$_2$Cu$_9$), respectively. The figures contain the complete experimental data that is provided as input to the CNNs (after unit conversion and wavelet transformation). The input thus consists of a three-channel scaleogram image obtained from $\{\mu_{z,\text{T} = 2\,\text{K}}, \chi_z(T), c_M(T)\}$ in the magnetic field and temperature ranges shown in the figure, where $z \equiv [001])$ direction. The experimental data is compared to theoretical results within the single-ion approximation using the values $\{\hat{x}_i\}$ predicted by the CNN. The numerical values of the Stevens coefficients obtained from the CNN are given in Table~\ref{tab:1} and~\ref{tab:2}. Ref.~\cite{MYERS199927} reports a value of $B^0_{2, \text{St.}} = 1.8 \pm 0.3$~K, which is close to the value $1.3$~K predicted by the CNN. Values for the other coefficients were not reported in Refs.~\cite{MYERS199927,PhysRevB.94.144434}.

To validate the CNN predictions, we use the coefficients $\{\hat{x}_i\}$ to calculate the thermodynamic observables (within the single-ion approximation). As shown in Fig.~\ref{fig:exp_res_combined}, we observe an overall very good agreement between the experimental data and the theoretical results for both compounds.  The MSEs are 0.074 for PrAgSb$_2$ and 0.085 for PrMg$_2$Cu$_9$, respectively. We note that due to the lack of data containing information about the magnetic anisotropy, the parameters $x_2, x_4$ (in tetragonal case) and $x_3$ (in hexagonal case) cannot be properly learned by the CNN. Its predictions are thus close to zero. As shown in Secs.~\ref{sec:results} and~\ref{subsec:CeAgSb2} this is not a limitation of our CNN approach and could be remedied by providing experimental magnetization and/or susceptibility data for a second direction, e.g., in the $ab$ plane. We note that Refs.~\cite{MYERS199927,PhysRevB.94.144434} contain results for in-plane directions, but do not determine the precise in-plane direction.

This justifies our approach to employ the single-ion approximation to describe the material properties, and demonstrates that the CNN has converged to a physically viable solution. In both cases, the network captures the initial slope of the magnetization [note that small $y$ axis scale in panel (d)]. It also correctly predicts both the Curie-Weiss slope of the inverse susceptibility at higher temperatures and the location of the Schottky peak in the specific heat. The latter is a direct indication that the energy level splitting between the ground and excited state has been successfully extracted from the data.

\begin{table}[tb]
    \centering
    \begin{tabular}{|c|c||c|c|c|}
    \hline
     & $x_i$ & $(k,q)$ & $\mathcal{B}^q_k$& $B^q_{k, \text{Stevens}}$  \\
    \hline
    $x_0$ & 57.9~K & -- & -- & -- \\
    \hline
    $x_1$ &0.575 & $(2,0)$ & 18.0~K & 5.7~K\\
    \hline
    $x_2$ &0.269 & $(4,0)$& 0.49~K & 0.05~K\\
    \hline
    $x_3$ & -- & $(6,6)$ & -- & -- \\
    \hline
    sign($x_4$) & $-1$ & $(6,0)$& 0.021~K & $1.3 \times 10^{-3}$~K\\
    \hline
    \end{tabular}
    \caption{Stevens parameters for PrMg$_2$Cu$_9$ obtained from CNN using $\{\mu_{z,T=2\,\text{K}}, \chi_z(T), c_M(T)\}$ from Ref.~\cite{PhysRevB.94.144434} as input data. Note that the CNN is unable to learn the coefficient $x_3$ since the experimental data set (and thus also the training data sets) does not include sufficient information about the magnetic anisotropy.}
    \label{tab:2}
\end{table}


\section{Summary and Outlook} 
\label{sec:summary_and_outlook}
To summarize, we present a deep ML algorithm for extracting CF Stevens parameters from thermodynamic observables of local-moment materials that can be treated within the single-ion approximation. We focus on rare-earth intermetallics and train a CNN on input data of magnetization, susceptibility, and specific heat. The training data is obtained from straightforward statistical mechanics calculations for different, randomly sampled, values of the Stevens parameters. To exploit the ability of CNNs in image recognition, we process the raw thermodynamic data using a wavelet transform and feed the resulting multi-channel scaleogram image to the network. The presented algorithm provides a convenient and powerful tool for extracting CF parameters from experimental data that avoids a tedious multiple parameter fitting procedure. We provide all programs necessary to run the algorithm and apply it to experimental situations as open source software~\cite{orth_berthusen_2020}.

The CNN provides an unbiased solution to the inverse problem of finding the Stevens parameters for a given set of thermodynamic observables. Depending on the type and amount of input data, this inverse problem can be ill-defined and allow for multiple solutions. Our study is an explicit test on the performance of CNNs on this well-known inverse physics problem of wide interest. We systematically investigate the performance of the algorithm for different site symmetries in the cubic, hexagonal and tetragonal crystal classes. The point groups we consider are experimentally relevant and allow for 2, 4 and 5 independent Stevens parameters, thus testing the CNN in cases of increasing complexity. We find that the CNN can accurately predict all Stevens coefficients if one provides magnetization data both along the easy-axis as well as within the easy-plane. The network performs equally well for integer and half-integer values of the total angular momentum $J$. Finally, we demonstrate that the algorithm also works well when applied to real experimental data of one Cerium system, CeAgSb$_2$~\cite{Takeuchi-PRB-2003}, and two Praseodymium compounds, PrAgSb$_2$~\cite{MYERS199927} and PrMg$_2$Cu$_9$~\cite{PhysRevB.94.144434}., which we obtain from the literature.

One promising future direction is to include correlation effects such as magnetic exchange interactions between different local moments within the modeling approach used to generate the training data. Magnetic exchange could, for example, be rather straightforwardly included via a molecular mean-field approach~\cite{Takeuchi-CeRhIn5-2001, Jobiliong-PRB-2005, Johnston-Unified_molecular_field-PRB-2015} (at the small cost of introducing an additional fit parameter describing the molecular field). Other interesting future directions are to apply more advanced ML techniques (generative adversarial networks ~\cite{goodfellow2014generative}, autoencoders) to lower symmetry CFs, to systematically investigate the stability of the algorithm with respect to input data noise, and to explore other choices of input observables, including low-symmetry magnetic field directions and direction averaged quantities. The latter would extend the applicability of the algorithm to polycrystalline materials.


\acknowledgments
The authors acknowledge valuable discussions with S.~L.~Bud'ko, P.~C.~Canfield, R.~J.~McQueeney, W.~R.~Meier and J.~S.~Van~Dyke.
This work was supported by the U.S. Department of Energy (DOE), Office of Science, Basic Energy Sciences, Division of Materials Sciences and Engineering. The research was performed at the Ames Laboratory, which is operated for the U.S. DOE by Iowa State University under Contract DE-AC02-07CH11358. M.S.S.~acknowledges support from the National Science Foundation under Grant No.~DMR-2002850.

\appendix
\section{Transformation between units used in experiment and training data}
\label{app:sec:units_transformation}
In this section, we describe how to transform from the commonly used experimental units in Refs.~\cite{MYERS199927,PhysRevB.94.144434} to the units used in the training data generation. It is important to first perform a transformation of units in the experimental data set before giving it to the CNN.
\subsection{Specific heat}
\label{app:ssub:specific_heat}
The SI units for the (magnetic) specific heat are $[c_M] = \text{J/(kg K)}$. One often also uses units of $[c_M] = \text{J/(mol K)}$. The training data contains $c_M/k_B$ per rare-earth ion, which is a dimensionless quantity that we denote by $c_\text{train} \equiv c/k_B$ per rare-earth ion. To transform experimental data $c_{\text{exp}}$ given in units of $\text{J/(mol K)}$ into the training data units, we need to perform
\begin{align}
 \frac{c_{\text{exp}}}{N_A k_B } = \frac{c_{\text{exp}}}{8.31445973} \; \widehat{=} \; c_{\text{train}}
\label{eq:specific_heat_unit_tf}
\end{align}
where $N_A$ is Avogadro's constant and $k_B$ is the Boltzmann constant. In other words, we divide the numerical values obtained from experimental plots by a factor of $8.31445973$ before feeding them into the CNN.

\subsection{Magnetization} 
\label{app:ssub:magnetization}
The training data uses the dimensionless quantity $\mu_\alpha/\mu_B$, which is the magnetic moment per rare-earth ion. This is identical to the units used in the experimental plots, which we can thus directly input into the CNN. We note that we transform from Oersted (auxiliary field units) to Tesla (magnetic field units) via $\mu_0 \times \; 1 \; \text{Oe} = 10^{-4} \; \text{T}$.

\subsection{Susceptibility} 
\label{app:ssub:susceptibility}
The magnetic susceptibility (or volume susceptibility) is experimentally obtained as $\chi = M/H$, where $M$ is the magnetization and $H$ is the auxiliary field. Its SI and cgs units are $[\chi] = 4 \pi (1) = 1 \text{emu}/\text{cm}^3$. Note that it is dimensionless in SI units. It is distinguished from the molar susceptibility $\chi_\text{mol}$ with units $[\chi_\text{mol}] = 4 \pi 10^{-6} \text{m}^3/\text{mol} = 1 \text{emu}/\text{mol}$. The two are related by $\chi = 4 \pi \times 10^{-6} \frac{\rho}{\mathcal{M}} \chi_\text{mol}$, where $\rho$ is the mass density, $\mathcal{M}$ is the molar mass.

The training data reads $\chi_\text{train} = \mu_\alpha/(\mu_B B)$, which is the ratio of the induced magnetic moment per rare-earth ion $\mu_a$ divided by the Bohr magneton $\mu_B$ and a small magnetic field $B = 10^{-4}$~T. Note that the result is independent of the value of $B$ as we ensure that we are in the linear regime of $\mu_a(B)$. This is related to the experimentally measured (volume) susceptibility via
$\chi_{\text{train}} =  \frac{V_{\text{uc}}}{N_R \mu_B \mu_0} \chi$, where $\chi$ is the magnetic susceptibility in SI units and $V_{\text{uc}}/N_R$ is an effective volume per rare-earth ion such that the magnetization $M_a = \mu_a/(V_{\text{uc}}/N_R)$. Here, $V_{\text{uc}}$ is the unit cell volume and $N_R$ is the number of rare-earth ions per unit cell.

Combining the two transformations discussed above leads to
\begin{align}
\chi_{\text{train}}  \, \widehat{=} \, 4 \pi \times 10^{-6} \frac{\rho}{\mathcal{M}} \frac{V_{\text{uc}}}{N_R} \frac{\chi_{\text{mol}} }{\mu_0 \mu_B} = 1.79053 \, \frac{N_{\text{f.u}}}{N_R} \chi_{\text{mol}}\,,
\label{eq:susc_exp_to_train}
\end{align}
where $N_{\text{f.u.}}$ is the number of formula units per unit cell. One thus needs to multiply the experimental data for $M/H$ (in emu/mol) by a factor of $1.79053 \frac{N_{\text{f.u}}}{N_R}$, before feeding it into the CNN.


\begin{thebibliography}{63}%
\makeatletter
\providecommand \@ifxundefined [1]{%
 \@ifx{#1\undefined}
}%
\providecommand \@ifnum [1]{%
 \ifnum #1\expandafter \@firstoftwo
 \else \expandafter \@secondoftwo
 \fi
}%
\providecommand \@ifx [1]{%
 \ifx #1\expandafter \@firstoftwo
 \else \expandafter \@secondoftwo
 \fi
}%
\providecommand \natexlab [1]{#1}%
\providecommand \enquote  [1]{``#1''}%
\providecommand \bibnamefont  [1]{#1}%
\providecommand \bibfnamefont [1]{#1}%
\providecommand \citenamefont [1]{#1}%
\providecommand \href@noop [0]{\@secondoftwo}%
\providecommand \href [0]{\begingroup \@sanitize@url \@href}%
\providecommand \@href[1]{\@@startlink{#1}\@@href}%
\providecommand \@@href[1]{\endgroup#1\@@endlink}%
\providecommand \@sanitize@url [0]{\catcode `\\12\catcode `\$12\catcode
  `\&12\catcode `\#12\catcode `\^12\catcode `\_12\catcode `\%12\relax}%
\providecommand \@@startlink[1]{}%
\providecommand \@@endlink[0]{}%
\providecommand \url  [0]{\begingroup\@sanitize@url \@url }%
\providecommand \@url [1]{\endgroup\@href {#1}{\urlprefix }}%
\providecommand \urlprefix  [0]{URL }%
\providecommand \Eprint [0]{\href }%
\providecommand \doibase [0]{https://doi.org/}%
\providecommand \selectlanguage [0]{\@gobble}%
\providecommand \bibinfo  [0]{\@secondoftwo}%
\providecommand \bibfield  [0]{\@secondoftwo}%
\providecommand \translation [1]{[#1]}%
\providecommand \BibitemOpen [0]{}%
\providecommand \bibitemStop [0]{}%
\providecommand \bibitemNoStop [0]{.\EOS\space}%
\providecommand \EOS [0]{\spacefactor3000\relax}%
\providecommand \BibitemShut  [1]{\csname bibitem#1\endcsname}%
\let\auto@bib@innerbib\@empty
\bibitem [{\citenamefont {Elliott}(1972)}]{elliottMagneticPropertiesRare1972}%
  \BibitemOpen
  \bibinfo {editor} {\bibfnamefont {R.}~\bibnamefont {Elliott}},\ ed.,\ \href
  {https://doi.org/10.1007/978-1-4757-5691-3} {\emph {\bibinfo {title}
  {Magnetic {{Properties}} of {{Rare Earth Metals}}}}}\ (\bibinfo  {publisher}
  {{Springer}},\ \bibinfo {year} {1972})\BibitemShut {NoStop}%
\bibitem [{\citenamefont {Fulde}\ and\ \citenamefont
  {Loewenhaupt}(1985)}]{fuldeMagneticExcitationsCrystalfield1985}%
  \BibitemOpen
  \bibfield  {author} {\bibinfo {author} {\bibfnamefont {P.}~\bibnamefont
  {Fulde}}\ and\ \bibinfo {author} {\bibfnamefont {M.}~\bibnamefont
  {Loewenhaupt}},\ }\bibfield  {title} {\bibinfo {title} {Magnetic excitations
  in crystal-field split 4f systems},\ }\href
  {https://doi.org/10.1080/00018738500101821} {\bibfield  {journal} {\bibinfo
  {journal} {Adv. Phys.}\ }\textbf {\bibinfo {volume} {34}},\ \bibinfo {pages}
  {589} (\bibinfo {year} {1985})}\BibitemShut {NoStop}%
\bibitem [{\citenamefont
  {Szytu{\l}a}(1991)}]{szytulaChapterMagneticProperties1991}%
  \BibitemOpen
  \bibfield  {author} {\bibinfo {author} {\bibfnamefont {A.}~\bibnamefont
  {Szytu{\l}a}},\ }\bibfield  {title} {\bibinfo {title} {Chapter 2 {{Magnetic}}
  properties of ternary intermetallic rare-earth compounds},\ }in\ \href
  {https://doi.org/10.1016/S1567-2719(05)80056-3} {\emph {\bibinfo {booktitle}
  {Handbook of {{Magnetic Materials}}}}},\ Vol.~\bibinfo {volume} {6}\
  (\bibinfo  {publisher} {{Elsevier}},\ \bibinfo {year} {1991})\ pp.\ \bibinfo
  {pages} {85--180}\BibitemShut {NoStop}%
\bibitem [{\citenamefont {Canfield}\ and\ \citenamefont
  {Bud'ko}(2016)}]{canfieldPreservedEntropyFragile2016}%
  \BibitemOpen
  \bibfield  {author} {\bibinfo {author} {\bibfnamefont {P.~C.}\ \bibnamefont
  {Canfield}}\ and\ \bibinfo {author} {\bibfnamefont {S.~L.}\ \bibnamefont
  {Bud'ko}},\ }\bibfield  {title} {\bibinfo {title} {Preserved entropy and
  fragile magnetism},\ }\href {https://doi.org/10.1088/0034-4885/79/8/084506}
  {\bibfield  {journal} {\bibinfo  {journal} {Rep. Prog. Phys.}\ }\textbf
  {\bibinfo {volume} {79}},\ \bibinfo {pages} {084506} (\bibinfo {year}
  {2016})}\BibitemShut {NoStop}%
\bibitem [{\citenamefont {Stevens}(1952)}]{stevensMatrixElementsOperator1952}%
  \BibitemOpen
  \bibfield  {author} {\bibinfo {author} {\bibfnamefont {K.~W.~H.}\
  \bibnamefont {Stevens}},\ }\bibfield  {title} {\bibinfo {title} {Matrix
  {{Elements}} and {{Operator Equivalents Connected}} with the {{Magnetic
  Properties}} of {{Rare Earth Ions}}},\ }\href
  {https://doi.org/10.1088/0370-1298/65/3/308} {\bibfield  {journal} {\bibinfo
  {journal} {Proc. Phys. Soc. A}\ }\textbf {\bibinfo {volume} {65}},\ \bibinfo
  {pages} {209} (\bibinfo {year} {1952})}\BibitemShut {NoStop}%
\bibitem [{\citenamefont {Bleaney}\ and\ \citenamefont
  {Stevens}(1953)}]{bleaneyParamagneticResonance1953}%
  \BibitemOpen
  \bibfield  {author} {\bibinfo {author} {\bibfnamefont {B.}~\bibnamefont
  {Bleaney}}\ and\ \bibinfo {author} {\bibfnamefont {K.~W.~H.}\ \bibnamefont
  {Stevens}},\ }\bibfield  {title} {\bibinfo {title} {Paramagnetic resonance},\
  }\href {https://doi.org/10.1088/0034-4885/16/1/304} {\bibfield  {journal}
  {\bibinfo  {journal} {Rep. Prog. Phys.}\ }\textbf {\bibinfo {volume} {16}},\
  \bibinfo {pages} {108} (\bibinfo {year} {1953})}\BibitemShut {NoStop}%
\bibitem [{\citenamefont {Altshuler}\ and\ \citenamefont
  {Kozyrev}(1964)}]{altshulerElectronParamagneticResonance1964}%
  \BibitemOpen
  \bibfield  {author} {\bibinfo {author} {\bibfnamefont {S.~A.}\ \bibnamefont
  {Altshuler}}\ and\ \bibinfo {author} {\bibfnamefont {B.~M.}\ \bibnamefont
  {Kozyrev}},\ }\href@noop {} {\emph {\bibinfo {title} {Electron Paramagnetic
  Resonance}}}\ (\bibinfo  {publisher} {{Academic Press}},\ \bibinfo {address}
  {{New York}},\ \bibinfo {year} {1964})\BibitemShut {NoStop}%
\bibitem [{\citenamefont
  {Wybourne}(1965)}]{wybourneSpectroscopicPropertiesRare1965}%
  \BibitemOpen
  \bibfield  {author} {\bibinfo {author} {\bibfnamefont {B.~G.}\ \bibnamefont
  {Wybourne}},\ }\href@noop {} {\emph {\bibinfo {title} {Spectroscopic
  Properties of Rare Earths}}}\ (\bibinfo  {publisher} {{John Wiley \& Sons}},\
  \bibinfo {address} {{New York, N.Y., USA}},\ \bibinfo {year}
  {1965})\BibitemShut {NoStop}%
\bibitem [{\citenamefont {Kuz'min}\ and\ \citenamefont
  {Tishin}(2007)}]{kuzminChapterThreeTheory2007}%
  \BibitemOpen
  \bibfield  {author} {\bibinfo {author} {\bibfnamefont {M.~D.}\ \bibnamefont
  {Kuz'min}}\ and\ \bibinfo {author} {\bibfnamefont {A.~M.}\ \bibnamefont
  {Tishin}},\ }\bibfield  {title} {\bibinfo {title} {Chapter {{Three Theory}}
  of {{Crystal}}-{{Field Effects}} in 3d-4f {{Intermetallic Compounds}}},\ }in\
  \href {https://doi.org/10.1016/S1567-2719(07)17003-7} {\emph {\bibinfo
  {booktitle} {Handbook of {{Magnetic Materials}}}}},\ Vol.~\bibinfo {volume}
  {17},\ \bibinfo {editor} {edited by\ \bibinfo {editor} {\bibfnamefont
  {K.~H.~J.}\ \bibnamefont {Buschow}}}\ (\bibinfo  {publisher} {{Elsevier}},\
  \bibinfo {year} {2007})\ pp.\ \bibinfo {pages} {149--233}\BibitemShut
  {NoStop}%
\bibitem [{\citenamefont {Fazekas}(1999)}]{fazekasLectureNotesElectron1999}%
  \BibitemOpen
  \bibfield  {author} {\bibinfo {author} {\bibfnamefont {P.}~\bibnamefont
  {Fazekas}},\ }\href@noop {} {\emph {\bibinfo {title} {Lecture Notes on
  Electron Correlation and Magnetism}}},\ \bibinfo {series} {Series in Modern
  Condensed Matter Physics}\ No.\ \bibinfo {number} {v. 5}\ (\bibinfo
  {publisher} {{World Scientific}},\ \bibinfo {address} {{Singapore ; River
  Edge, N.J}},\ \bibinfo {year} {1999})\BibitemShut {NoStop}%
\bibitem [{\citenamefont {Cox}(1987)}]{coxQuadrupolarKondoEffect1987}%
  \BibitemOpen
  \bibfield  {author} {\bibinfo {author} {\bibfnamefont {D.~L.}\ \bibnamefont
  {Cox}},\ }\bibfield  {title} {\bibinfo {title} {Quadrupolar {{Kondo}} effect
  in uranium heavy-electron materials?},\ }\href
  {https://doi.org/10.1103/PhysRevLett.59.1240} {\bibfield  {journal} {\bibinfo
   {journal} {Phys. Rev. Lett.}\ }\textbf {\bibinfo {volume} {59}},\ \bibinfo
  {pages} {1240} (\bibinfo {year} {1987})}\BibitemShut {NoStop}%
\bibitem [{\citenamefont {Cox}\ and\ \citenamefont
  {Zawadowski}(1998)}]{coxExoticKondoEffects1998}%
  \BibitemOpen
  \bibfield  {author} {\bibinfo {author} {\bibfnamefont {D.~L.}\ \bibnamefont
  {Cox}}\ and\ \bibinfo {author} {\bibfnamefont {A.}~\bibnamefont
  {Zawadowski}},\ }\bibfield  {title} {\bibinfo {title} {Exotic {{Kondo}}
  effects in metals: {{Magnetic}} ions in a crystalline electric field and
  tunnelling centres},\ }\href {https://doi.org/10.1080/000187398243500}
  {\bibfield  {journal} {\bibinfo  {journal} {Advances in Physics}\ }\textbf
  {\bibinfo {volume} {47}},\ \bibinfo {pages} {599} (\bibinfo {year}
  {1998})}\BibitemShut {NoStop}%
\bibitem [{\citenamefont {Levy}\ and\ \citenamefont
  {Zhang}(1989)}]{Levy-PRL-1989}%
  \BibitemOpen
  \bibfield  {author} {\bibinfo {author} {\bibfnamefont {P.~M.}\ \bibnamefont
  {Levy}}\ and\ \bibinfo {author} {\bibfnamefont {S.}~\bibnamefont {Zhang}},\
  }\bibfield  {title} {\bibinfo {title} {Crystal-field splitting in kondo
  systems},\ }\href {https://doi.org/10.1103/PhysRevLett.62.78} {\bibfield
  {journal} {\bibinfo  {journal} {Phys. Rev. Lett.}\ }\textbf {\bibinfo
  {volume} {62}},\ \bibinfo {pages} {78} (\bibinfo {year} {1989})}\BibitemShut
  {NoStop}%
\bibitem [{\citenamefont {Ikeda}\ and\ \citenamefont
  {Miyake}(1996)}]{ikedaTheoryAnisotropicSemiconductor1996}%
  \BibitemOpen
  \bibfield  {author} {\bibinfo {author} {\bibfnamefont {H.}~\bibnamefont
  {Ikeda}}\ and\ \bibinfo {author} {\bibfnamefont {K.}~\bibnamefont {Miyake}},\
  }\bibfield  {title} {\bibinfo {title} {A {{Theory}} of {{Anisotropic
  Semiconductor}} of {{Heavy Fermions}}},\ }\href
  {https://doi.org/10.1143/JPSJ.65.1769} {\bibfield  {journal} {\bibinfo
  {journal} {J. Phys. Soc. Jpn.}\ }\textbf {\bibinfo {volume} {65}},\ \bibinfo
  {pages} {1769} (\bibinfo {year} {1996})}\BibitemShut {NoStop}%
\bibitem [{\citenamefont {Anders}\ and\ \citenamefont
  {Pruschke}(2006)}]{Anders-PRL-2006}%
  \BibitemOpen
  \bibfield  {author} {\bibinfo {author} {\bibfnamefont {F.~B.}\ \bibnamefont
  {Anders}}\ and\ \bibinfo {author} {\bibfnamefont {T.}~\bibnamefont
  {Pruschke}},\ }\bibfield  {title} {\bibinfo {title} {Can competition between
  the crystal field and the kondo effect cause non-fermi-liquid-like
  behavior?},\ }\href {https://doi.org/10.1103/PhysRevLett.96.086404}
  {\bibfield  {journal} {\bibinfo  {journal} {Phys. Rev. Lett.}\ }\textbf
  {\bibinfo {volume} {96}},\ \bibinfo {pages} {086404} (\bibinfo {year}
  {2006})}\BibitemShut {NoStop}%
\bibitem [{\citenamefont {Peyker}\ \emph {et~al.}(2009)\citenamefont {Peyker},
  \citenamefont {Gold}, \citenamefont {Scheidt}, \citenamefont {Scherer},
  \citenamefont {Donath}, \citenamefont {Gegenwart}, \citenamefont {Mayr},
  \citenamefont {Unruh}, \citenamefont {Eyert}, \citenamefont {Bauer},\ and\
  \citenamefont {Michor}}]{Peyker_2009}%
  \BibitemOpen
  \bibfield  {author} {\bibinfo {author} {\bibfnamefont {L.}~\bibnamefont
  {Peyker}}, \bibinfo {author} {\bibfnamefont {C.}~\bibnamefont {Gold}},
  \bibinfo {author} {\bibfnamefont {E.-W.}\ \bibnamefont {Scheidt}}, \bibinfo
  {author} {\bibfnamefont {W.}~\bibnamefont {Scherer}}, \bibinfo {author}
  {\bibfnamefont {J.~G.}\ \bibnamefont {Donath}}, \bibinfo {author}
  {\bibfnamefont {P.}~\bibnamefont {Gegenwart}}, \bibinfo {author}
  {\bibfnamefont {F.}~\bibnamefont {Mayr}}, \bibinfo {author} {\bibfnamefont
  {T.}~\bibnamefont {Unruh}}, \bibinfo {author} {\bibfnamefont
  {V.}~\bibnamefont {Eyert}}, \bibinfo {author} {\bibfnamefont
  {E.}~\bibnamefont {Bauer}},\ and\ \bibinfo {author} {\bibfnamefont
  {H.}~\bibnamefont {Michor}},\ }\bibfield  {title} {\bibinfo {title}
  {Evolution of quantum criticality in {CeNi}9-{xCuxGe}4},\ }\href
  {https://doi.org/10.1088/0953-8984/21/23/235604} {\bibfield  {journal}
  {\bibinfo  {journal} {J. Phys.: Condens. Matter}\ }\textbf {\bibinfo {volume}
  {21}},\ \bibinfo {pages} {235604} (\bibinfo {year} {2009})}\BibitemShut
  {NoStop}%
\bibitem [{\citenamefont {Dzero}\ \emph {et~al.}(2010)\citenamefont {Dzero},
  \citenamefont {Sun}, \citenamefont {Galitski},\ and\ \citenamefont
  {Coleman}}]{Dzero-PRL-2010}%
  \BibitemOpen
  \bibfield  {author} {\bibinfo {author} {\bibfnamefont {M.}~\bibnamefont
  {Dzero}}, \bibinfo {author} {\bibfnamefont {K.}~\bibnamefont {Sun}}, \bibinfo
  {author} {\bibfnamefont {V.}~\bibnamefont {Galitski}},\ and\ \bibinfo
  {author} {\bibfnamefont {P.}~\bibnamefont {Coleman}},\ }\bibfield  {title}
  {\bibinfo {title} {Topological kondo insulators},\ }\href
  {https://doi.org/10.1103/PhysRevLett.104.106408} {\bibfield  {journal}
  {\bibinfo  {journal} {Phys. Rev. Lett.}\ }\textbf {\bibinfo {volume} {104}},\
  \bibinfo {pages} {106408} (\bibinfo {year} {2010})}\BibitemShut {NoStop}%
\bibitem [{\citenamefont {Romero}\ \emph {et~al.}(2013)\citenamefont {Romero},
  \citenamefont {Aligia}, \citenamefont {Sereni},\ and\ \citenamefont
  {Nieva}}]{Romero_2013}%
  \BibitemOpen
  \bibfield  {author} {\bibinfo {author} {\bibfnamefont {M.~A.}\ \bibnamefont
  {Romero}}, \bibinfo {author} {\bibfnamefont {A.~A.}\ \bibnamefont {Aligia}},
  \bibinfo {author} {\bibfnamefont {J.~G.}\ \bibnamefont {Sereni}},\ and\
  \bibinfo {author} {\bibfnamefont {G.}~\bibnamefont {Nieva}},\ }\bibfield
  {title} {\bibinfo {title} {Interpretation of experimental results on
  {{K}}ondo systems with crystal field},\ }\href
  {https://doi.org/10.1088/0953-8984/26/2/025602} {\bibfield  {journal}
  {\bibinfo  {journal} {J. Phys.: Condens. Matter}\ }\textbf {\bibinfo {volume}
  {26}},\ \bibinfo {pages} {025602} (\bibinfo {year} {2013})}\BibitemShut
  {NoStop}%
\bibitem [{\citenamefont {Desgranges}(2014)}]{Desgranges-Physica_B-2014}%
  \BibitemOpen
  \bibfield  {author} {\bibinfo {author} {\bibfnamefont {H.-U.}\ \bibnamefont
  {Desgranges}},\ }\bibfield  {title} {\bibinfo {title} {Crystal fields and
  kondo effect: Specific heat for cerium compounds},\ }\href
  {https://doi.org/https://doi.org/10.1016/j.physb.2014.07.077} {\bibfield
  {journal} {\bibinfo  {journal} {Physica B: Condensed Matter}\ }\textbf
  {\bibinfo {volume} {454}},\ \bibinfo {pages} {135} (\bibinfo {year}
  {2014})}\BibitemShut {NoStop}%
\bibitem [{\citenamefont {Chandra}\ \emph {et~al.}(2013)\citenamefont
  {Chandra}, \citenamefont {Coleman},\ and\ \citenamefont
  {Flint}}]{chandraHastaticOrderHeavyfermion2013}%
  \BibitemOpen
  \bibfield  {author} {\bibinfo {author} {\bibfnamefont {P.}~\bibnamefont
  {Chandra}}, \bibinfo {author} {\bibfnamefont {P.}~\bibnamefont {Coleman}},\
  and\ \bibinfo {author} {\bibfnamefont {R.}~\bibnamefont {Flint}},\ }\bibfield
   {title} {\bibinfo {title} {Hastatic order in the heavy-fermion compound
  {{URu2Si2}}},\ }\href {https://doi.org/10.1038/nature11820} {\bibfield
  {journal} {\bibinfo  {journal} {Nature}\ }\textbf {\bibinfo {volume} {493}},\
  \bibinfo {pages} {621} (\bibinfo {year} {2013})}\BibitemShut {NoStop}%
\bibitem [{\citenamefont {Van~Dyke}\ \emph {et~al.}(2019)\citenamefont
  {Van~Dyke}, \citenamefont {Zhang},\ and\ \citenamefont
  {Flint}}]{vandykeFieldinducedFerrohastaticPhase2019}%
  \BibitemOpen
  \bibfield  {author} {\bibinfo {author} {\bibfnamefont {J.~S.}\ \bibnamefont
  {Van~Dyke}}, \bibinfo {author} {\bibfnamefont {G.}~\bibnamefont {Zhang}},\
  and\ \bibinfo {author} {\bibfnamefont {R.}~\bibnamefont {Flint}},\ }\bibfield
   {title} {\bibinfo {title} {Field-induced ferrohastatic phase in cubic
  non-{{Kramers}} doublet systems},\ }\href
  {https://doi.org/10.1103/PhysRevB.100.205122} {\bibfield  {journal} {\bibinfo
   {journal} {Phys. Rev. B}\ }\textbf {\bibinfo {volume} {100}},\ \bibinfo
  {pages} {205122} (\bibinfo {year} {2019})}\BibitemShut {NoStop}%
\bibitem [{\citenamefont {Bethe}(1929)}]{betheTermaufspaltungKristallen1929}%
  \BibitemOpen
  \bibfield  {author} {\bibinfo {author} {\bibfnamefont {H.}~\bibnamefont
  {Bethe}},\ }\bibfield  {title} {\bibinfo {title} {{Termaufspaltung in
  Kristallen}},\ }\href {https://doi.org/10.1002/andp.19293950202} {\bibfield
  {journal} {\bibinfo  {journal} {Annalen der Physik}\ }\textbf {\bibinfo
  {volume} {395}},\ \bibinfo {pages} {133} (\bibinfo {year}
  {1929})}\BibitemShut {NoStop}%
\bibitem [{\citenamefont {Lea}\ \emph {et~al.}(1962)\citenamefont {Lea},
  \citenamefont {Leask},\ and\ \citenamefont
  {Wolf}}]{leaRaisingAngularMomentum1962}%
  \BibitemOpen
  \bibfield  {author} {\bibinfo {author} {\bibfnamefont {K.}~\bibnamefont
  {Lea}}, \bibinfo {author} {\bibfnamefont {M.}~\bibnamefont {Leask}},\ and\
  \bibinfo {author} {\bibfnamefont {W.}~\bibnamefont {Wolf}},\ }\bibfield
  {title} {\bibinfo {title} {The raising of angular momentum degeneracy of
  f-{{Electron}} terms by cubic crystal fields},\ }\href
  {https://doi.org/10.1016/0022-3697(62)90192-0} {\bibfield  {journal}
  {\bibinfo  {journal} {J. Phys. Chem. Solids}\ }\textbf {\bibinfo {volume}
  {23}},\ \bibinfo {pages} {1381} (\bibinfo {year} {1962})}\BibitemShut
  {NoStop}%
\bibitem [{\citenamefont {Walter}(1984)}]{walterTreatingCrystalField1984}%
  \BibitemOpen
  \bibfield  {author} {\bibinfo {author} {\bibfnamefont {U.}~\bibnamefont
  {Walter}},\ }\bibfield  {title} {\bibinfo {title} {Treating crystal field
  parameters in lower than cubic symmetries},\ }\href
  {https://doi.org/10.1016/0022-3697(84)90147-1} {\bibfield  {journal}
  {\bibinfo  {journal} {J. Phys. Chem. Solids}\ }\textbf {\bibinfo {volume}
  {45}},\ \bibinfo {pages} {401} (\bibinfo {year} {1984})}\BibitemShut
  {NoStop}%
\bibitem [{\citenamefont
  {Richter}(2001)}]{richterChapterDensityFunctional2001}%
  \BibitemOpen
  \bibfield  {author} {\bibinfo {author} {\bibfnamefont {M.}~\bibnamefont
  {Richter}},\ }\bibfield  {title} {\bibinfo {title} {Chapter 2 {{Density}}
  functional theory applied to 4f and 5f elements and metallic compounds},\
  }in\ \href {https://doi.org/10.1016/S1567-2719(01)13006-4} {\emph {\bibinfo
  {booktitle} {Handbook of {{Magnetic Materials}}}}},\ Vol.~\bibinfo {volume}
  {13}\ (\bibinfo  {publisher} {{Elsevier}},\ \bibinfo {year} {2001})\ pp.\
  \bibinfo {pages} {87--228}\BibitemShut {NoStop}%
\bibitem [{\citenamefont {Loewenhaupt}\ and\ \citenamefont
  {Fischer}(1993)}]{loewenhauptChapterNeutronScattering1993}%
  \BibitemOpen
  \bibfield  {author} {\bibinfo {author} {\bibfnamefont {M.}~\bibnamefont
  {Loewenhaupt}}\ and\ \bibinfo {author} {\bibfnamefont {K.~H.}\ \bibnamefont
  {Fischer}},\ }\bibfield  {title} {\bibinfo {title} {Chapter 6 {{Neutron}}
  scattering on heavy fermion and valence fluctuation 4f-systems},\ }in\ \href
  {https://doi.org/10.1016/S1567-2719(05)80047-2} {\emph {\bibinfo {booktitle}
  {Handbook of {{Magnetic Materials}}}}},\ Vol.~\bibinfo {volume} {7}\
  (\bibinfo  {publisher} {{Elsevier}},\ \bibinfo {year} {1993})\ pp.\ \bibinfo
  {pages} {503--608}\BibitemShut {NoStop}%
\bibitem [{\citenamefont {Moze}(1998)}]{mozeChapterCrystalField1998}%
  \BibitemOpen
  \bibfield  {author} {\bibinfo {author} {\bibfnamefont {O.}~\bibnamefont
  {Moze}},\ }\bibfield  {title} {\bibinfo {title} {Chapter 4 {{Crystal}} field
  effects in intermetallic compounds studied by inelastic neutron scattering},\
  }in\ \href {https://doi.org/10.1016/S1567-2719(98)11008-9} {\emph {\bibinfo
  {booktitle} {Handbook of {{Magnetic Materials}}}}},\ Vol.~\bibinfo {volume}
  {11}\ (\bibinfo  {publisher} {{Elsevier}},\ \bibinfo {year} {1998})\ pp.\
  \bibinfo {pages} {493--624}\BibitemShut {NoStop}%
\bibitem [{\citenamefont {Myers}\ \emph {et~al.}(1999)\citenamefont {Myers},
  \citenamefont {Bud'ko}, \citenamefont {Fisher}, \citenamefont {Islam},
  \citenamefont {Kleinke}, \citenamefont {Lacerda},\ and\ \citenamefont
  {Canfield}}]{MYERS199927}%
  \BibitemOpen
  \bibfield  {author} {\bibinfo {author} {\bibfnamefont {K.}~\bibnamefont
  {Myers}}, \bibinfo {author} {\bibfnamefont {S.}~\bibnamefont {Bud'ko}},
  \bibinfo {author} {\bibfnamefont {I.}~\bibnamefont {Fisher}}, \bibinfo
  {author} {\bibfnamefont {Z.}~\bibnamefont {Islam}}, \bibinfo {author}
  {\bibfnamefont {H.}~\bibnamefont {Kleinke}}, \bibinfo {author} {\bibfnamefont
  {A.}~\bibnamefont {Lacerda}},\ and\ \bibinfo {author} {\bibfnamefont
  {P.}~\bibnamefont {Canfield}},\ }\bibfield  {title} {\bibinfo {title}
  {Systematic study of anisotropic transport and magnetic properties of
  {{RAgSb2}} ({{R}}={{Y}}, {{La}}\textendash{{Nd}}, sm,
  {{Gd}}\textendash{{Tm}})},\ }\href
  {https://doi.org/10.1016/S0304-8853(99)00472-2} {\bibfield  {journal}
  {\bibinfo  {journal} {J. Magn. Magn. Mater}\ }\textbf {\bibinfo {volume}
  {205}},\ \bibinfo {pages} {27} (\bibinfo {year} {1999})}\BibitemShut
  {NoStop}%
\bibitem [{\citenamefont {Bud'ko}\ \emph {et~al.}(1999)\citenamefont {Bud'ko},
  \citenamefont {Islam}, \citenamefont {Wiener}, \citenamefont {Fisher},
  \citenamefont {Lacerda},\ and\ \citenamefont
  {Canfield}}]{budkoAnisotropyMetamagnetismRNi2Ge21999}%
  \BibitemOpen
  \bibfield  {author} {\bibinfo {author} {\bibfnamefont {S.~L.}\ \bibnamefont
  {Bud'ko}}, \bibinfo {author} {\bibfnamefont {Z.}~\bibnamefont {Islam}},
  \bibinfo {author} {\bibfnamefont {T.~A.}\ \bibnamefont {Wiener}}, \bibinfo
  {author} {\bibfnamefont {I.~R.}\ \bibnamefont {Fisher}}, \bibinfo {author}
  {\bibfnamefont {A.~H.}\ \bibnamefont {Lacerda}},\ and\ \bibinfo {author}
  {\bibfnamefont {P.~C.}\ \bibnamefont {Canfield}},\ }\bibfield  {title}
  {\bibinfo {title} {Anisotropy and metamagnetism in the {{RNi2Ge2}}
  ({{R}}={{Y}}, {{La}}\textendash{{Nd}}, {{Sm}}\textendash{{Lu}}) series},\
  }\href {https://doi.org/10.1016/S0304-8853(99)00486-2} {\bibfield  {journal}
  {\bibinfo  {journal} {J. Magn. Magn. Mater}\ }\textbf {\bibinfo {volume}
  {205}},\ \bibinfo {pages} {53} (\bibinfo {year} {1999})}\BibitemShut
  {NoStop}%
\bibitem [{\citenamefont {Kong}\ \emph {et~al.}(2016)\citenamefont {Kong},
  \citenamefont {Meier}, \citenamefont {Lin}, \citenamefont {Saunders},
  \citenamefont {Bud'ko}, \citenamefont {Flint},\ and\ \citenamefont
  {Canfield}}]{PhysRevB.94.144434}%
  \BibitemOpen
  \bibfield  {author} {\bibinfo {author} {\bibfnamefont {T.}~\bibnamefont
  {Kong}}, \bibinfo {author} {\bibfnamefont {W.~R.}\ \bibnamefont {Meier}},
  \bibinfo {author} {\bibfnamefont {Q.}~\bibnamefont {Lin}}, \bibinfo {author}
  {\bibfnamefont {S.~M.}\ \bibnamefont {Saunders}}, \bibinfo {author}
  {\bibfnamefont {S.~L.}\ \bibnamefont {Bud'ko}}, \bibinfo {author}
  {\bibfnamefont {R.}~\bibnamefont {Flint}},\ and\ \bibinfo {author}
  {\bibfnamefont {P.~C.}\ \bibnamefont {Canfield}},\ }\bibfield  {title}
  {\bibinfo {title} {Physical properties of single crystalline
  {{RMg}}{$_{2}$}{{Cu}}{$_9$} ({{R}}={{Y}},{{Ce}}-{{Nd}},{{Gd}}-{{Dy}},{{Yb}})
  and the search for in-plane magnetic anisotropy in hexagonal systems},\
  }\href {https://doi.org/10.1103/PhysRevB.94.144434} {\bibfield  {journal}
  {\bibinfo  {journal} {Phys. Rev. B}\ }\textbf {\bibinfo {volume} {94}},\
  \bibinfo {pages} {144434} (\bibinfo {year} {2016})}\BibitemShut {NoStop}%
\bibitem [{\citenamefont {Geilhufe}\ and\ \citenamefont
  {Hergert}(2018)}]{gtpack1}%
  \BibitemOpen
  \bibfield  {author} {\bibinfo {author} {\bibfnamefont {R.~M.}\ \bibnamefont
  {Geilhufe}}\ and\ \bibinfo {author} {\bibfnamefont {W.}~\bibnamefont
  {Hergert}},\ }\bibfield  {title} {\bibinfo {title} {{{GTPack}}: {{A}}
  mathematica group theory package for application in solid-state physics and
  photonics},\ }\href {https://doi.org/10.3389/fphy.2018.00086} {\bibfield
  {journal} {\bibinfo  {journal} {Frontiers in Physics}\ }\textbf {\bibinfo
  {volume} {6}},\ \bibinfo {pages} {86} (\bibinfo {year} {2018})}\BibitemShut
  {NoStop}%
\bibitem [{\citenamefont {Hergert}\ and\ \citenamefont
  {Geilhufe}(2017)}]{gtpack2}%
  \BibitemOpen
  \bibfield  {author} {\bibinfo {author} {\bibfnamefont {W.}~\bibnamefont
  {Hergert}}\ and\ \bibinfo {author} {\bibfnamefont {R.~M.}\ \bibnamefont
  {Geilhufe}},\ }\href@noop {} {\emph {\bibinfo {title} {Group Theory in Solid
  State Physics and Photonics: Problem Solving with Mathematica}}}\ (\bibinfo
  {publisher} {{Wiley-VCH}},\ \bibinfo {address} {{Weinheim, Germany}},\
  \bibinfo {year} {2017})\BibitemShut {NoStop}%
\bibitem [{\citenamefont {Engl}\ \emph {et~al.}(1996)\citenamefont {Engl},
  \citenamefont {Hanke},\ and\ \citenamefont
  {Neubauer}}]{engl1996regularization}%
  \BibitemOpen
  \bibfield  {author} {\bibinfo {author} {\bibfnamefont {H.~W.}\ \bibnamefont
  {Engl}}, \bibinfo {author} {\bibfnamefont {M.}~\bibnamefont {Hanke}},\ and\
  \bibinfo {author} {\bibfnamefont {A.}~\bibnamefont {Neubauer}},\ }\href@noop
  {} {\emph {\bibinfo {title} {Regularization of Inverse Problems}}},\ Vol.\
  \bibinfo {volume} {375}\ (\bibinfo  {publisher} {{Springer Science \&
  Business Media}},\ \bibinfo {year} {1996})\BibitemShut {NoStop}%
\bibitem [{\citenamefont {Carleo}\ \emph {et~al.}(2019)\citenamefont {Carleo},
  \citenamefont {Cirac}, \citenamefont {Cranmer}, \citenamefont {Daudet},
  \citenamefont {Schuld}, \citenamefont {Tishby}, \citenamefont
  {{Vogt-Maranto}},\ and\ \citenamefont {Zdeborov{\'a}}}]{RMPMachineLearning}%
  \BibitemOpen
  \bibfield  {author} {\bibinfo {author} {\bibfnamefont {G.}~\bibnamefont
  {Carleo}}, \bibinfo {author} {\bibfnamefont {I.}~\bibnamefont {Cirac}},
  \bibinfo {author} {\bibfnamefont {K.}~\bibnamefont {Cranmer}}, \bibinfo
  {author} {\bibfnamefont {L.}~\bibnamefont {Daudet}}, \bibinfo {author}
  {\bibfnamefont {M.}~\bibnamefont {Schuld}}, \bibinfo {author} {\bibfnamefont
  {N.}~\bibnamefont {Tishby}}, \bibinfo {author} {\bibfnamefont
  {L.}~\bibnamefont {{Vogt-Maranto}}},\ and\ \bibinfo {author} {\bibfnamefont
  {L.}~\bibnamefont {Zdeborov{\'a}}},\ }\bibfield  {title} {\bibinfo {title}
  {Machine learning and the physical sciences},\ }\href
  {https://doi.org/10.1103/RevModPhys.91.045002} {\bibfield  {journal}
  {\bibinfo  {journal} {Rev. Mod. Phys.}\ }\textbf {\bibinfo {volume} {91}},\
  \bibinfo {pages} {045002} (\bibinfo {year} {2019})}\BibitemShut {NoStop}%
\bibitem [{\citenamefont {Mehta}\ \emph {et~al.}(2019)\citenamefont {Mehta},
  \citenamefont {Bukov}, \citenamefont {Wang}, \citenamefont {Day},
  \citenamefont {Richardson}, \citenamefont {Fisher},\ and\ \citenamefont
  {Schwab}}]{MEHTA20191}%
  \BibitemOpen
  \bibfield  {author} {\bibinfo {author} {\bibfnamefont {P.}~\bibnamefont
  {Mehta}}, \bibinfo {author} {\bibfnamefont {M.}~\bibnamefont {Bukov}},
  \bibinfo {author} {\bibfnamefont {C.-H.}\ \bibnamefont {Wang}}, \bibinfo
  {author} {\bibfnamefont {A.~G.}\ \bibnamefont {Day}}, \bibinfo {author}
  {\bibfnamefont {C.}~\bibnamefont {Richardson}}, \bibinfo {author}
  {\bibfnamefont {C.~K.}\ \bibnamefont {Fisher}},\ and\ \bibinfo {author}
  {\bibfnamefont {D.~J.}\ \bibnamefont {Schwab}},\ }\bibfield  {title}
  {\bibinfo {title} {A high-bias, low-variance introduction to {{Machine
  Learning}} for physicists},\ }\href
  {https://doi.org/10.1016/j.physrep.2019.03.001} {\bibfield  {journal}
  {\bibinfo  {journal} {Phys. Rep.}\ }\textbf {\bibinfo {volume} {810}},\
  \bibinfo {pages} {1} (\bibinfo {year} {2019})}\BibitemShut {NoStop}%
\bibitem [{\citenamefont {Dunjko}\ and\ \citenamefont
  {Briegel}(2018)}]{Dunjko_2018}%
  \BibitemOpen
  \bibfield  {author} {\bibinfo {author} {\bibfnamefont {V.}~\bibnamefont
  {Dunjko}}\ and\ \bibinfo {author} {\bibfnamefont {H.~J.}\ \bibnamefont
  {Briegel}},\ }\bibfield  {title} {\bibinfo {title} {Machine learning \&
  artificial intelligence in the quantum domain: A review of recent progress},\
  }\href {https://doi.org/10.1088/1361-6633/aab406} {\bibfield  {journal}
  {\bibinfo  {journal} {Rep. Prog. Phys.}\ }\textbf {\bibinfo {volume} {81}},\
  \bibinfo {pages} {074001} (\bibinfo {year} {2018})}\BibitemShut {NoStop}%
\bibitem [{\citenamefont {Arsenault}\ \emph {et~al.}(2017)\citenamefont
  {Arsenault}, \citenamefont {Neuberg}, \citenamefont {Hannah},\ and\
  \citenamefont {Millis}}]{Arsenault_2017}%
  \BibitemOpen
  \bibfield  {author} {\bibinfo {author} {\bibfnamefont {L.-F.}\ \bibnamefont
  {Arsenault}}, \bibinfo {author} {\bibfnamefont {R.}~\bibnamefont {Neuberg}},
  \bibinfo {author} {\bibfnamefont {L.~A.}\ \bibnamefont {Hannah}},\ and\
  \bibinfo {author} {\bibfnamefont {A.~J.}\ \bibnamefont {Millis}},\ }\bibfield
   {title} {\bibinfo {title} {Projected regression method for solving
  {{Fredholm}} integral equations arising in the analytic continuation problem
  of quantum physics},\ }\href {https://doi.org/10.1088/1361-6420/aa8d93}
  {\bibfield  {journal} {\bibinfo  {journal} {Inverse Problems}\ }\textbf
  {\bibinfo {volume} {33}},\ \bibinfo {pages} {115007} (\bibinfo {year}
  {2017})}\BibitemShut {NoStop}%
\bibitem [{\citenamefont {Fournier}\ \emph {et~al.}(2020)\citenamefont
  {Fournier}, \citenamefont {Wang}, \citenamefont {Yazyev},\ and\ \citenamefont
  {Wu}}]{PhysRevLett.124.056401}%
  \BibitemOpen
  \bibfield  {author} {\bibinfo {author} {\bibfnamefont {R.}~\bibnamefont
  {Fournier}}, \bibinfo {author} {\bibfnamefont {L.}~\bibnamefont {Wang}},
  \bibinfo {author} {\bibfnamefont {O.~V.}\ \bibnamefont {Yazyev}},\ and\
  \bibinfo {author} {\bibfnamefont {Q.}~\bibnamefont {Wu}},\ }\bibfield
  {title} {\bibinfo {title} {Artificial neural network approach to the analytic
  continuation problem},\ }\href
  {https://doi.org/10.1103/PhysRevLett.124.056401} {\bibfield  {journal}
  {\bibinfo  {journal} {Phys. Rev. Lett.}\ }\textbf {\bibinfo {volume} {124}},\
  \bibinfo {pages} {056401} (\bibinfo {year} {2020})}\BibitemShut {NoStop}%
\bibitem [{\citenamefont {Hanakata}\ \emph {et~al.}(2020)\citenamefont
  {Hanakata}, \citenamefont {Cubuk}, \citenamefont {Campbell},\ and\
  \citenamefont {Park}}]{SupervisedAutoencoder}%
  \BibitemOpen
  \bibfield  {author} {\bibinfo {author} {\bibfnamefont {P.~Z.}\ \bibnamefont
  {Hanakata}}, \bibinfo {author} {\bibfnamefont {E.~D.}\ \bibnamefont {Cubuk}},
  \bibinfo {author} {\bibfnamefont {D.~K.}\ \bibnamefont {Campbell}},\ and\
  \bibinfo {author} {\bibfnamefont {H.~S.}\ \bibnamefont {Park}},\ }\href@noop
  {} {\bibinfo {title} {Forward and inverse design of kirigami via supervised
  autoencoder}} (\bibinfo {year} {2020}),\ \Eprint
  {https://arxiv.org/abs/2008.05298} {arXiv:2008.05298} \BibitemShut {NoStop}%
\bibitem [{\citenamefont {Teoh}\ \emph {et~al.}(2020)\citenamefont {Teoh},
  \citenamefont {Drygala}, \citenamefont {Melko},\ and\ \citenamefont
  {Islam}}]{InverseDesignLaser}%
  \BibitemOpen
  \bibfield  {author} {\bibinfo {author} {\bibfnamefont {Y.~H.}\ \bibnamefont
  {Teoh}}, \bibinfo {author} {\bibfnamefont {M.}~\bibnamefont {Drygala}},
  \bibinfo {author} {\bibfnamefont {R.~G.}\ \bibnamefont {Melko}},\ and\
  \bibinfo {author} {\bibfnamefont {R.}~\bibnamefont {Islam}},\ }\bibfield
  {title} {\bibinfo {title} {Machine learning design of a trapped-ion quantum
  spin simulator},\ }\href {https://doi.org/10.1088/2058-9565/ab657a}
  {\bibfield  {journal} {\bibinfo  {journal} {Quantum Sci. Technol.}\ }\textbf
  {\bibinfo {volume} {5}},\ \bibinfo {pages} {024001} (\bibinfo {year}
  {2020})}\BibitemShut {NoStop}%
\bibitem [{\citenamefont {Laanait}\ \emph {et~al.}(2019)\citenamefont
  {Laanait}, \citenamefont {Romero}, \citenamefont {Yin}, \citenamefont
  {Young}, \citenamefont {Treichler}, \citenamefont {Starchenko}, \citenamefont
  {Borisevich}, \citenamefont {Sergeev},\ and\ \citenamefont
  {Matheson}}]{2019arXiv190911150L}%
  \BibitemOpen
  \bibfield  {author} {\bibinfo {author} {\bibfnamefont {N.}~\bibnamefont
  {Laanait}}, \bibinfo {author} {\bibfnamefont {J.}~\bibnamefont {Romero}},
  \bibinfo {author} {\bibfnamefont {J.}~\bibnamefont {Yin}}, \bibinfo {author}
  {\bibfnamefont {M.~T.}\ \bibnamefont {Young}}, \bibinfo {author}
  {\bibfnamefont {S.}~\bibnamefont {Treichler}}, \bibinfo {author}
  {\bibfnamefont {V.}~\bibnamefont {Starchenko}}, \bibinfo {author}
  {\bibfnamefont {A.}~\bibnamefont {Borisevich}}, \bibinfo {author}
  {\bibfnamefont {A.}~\bibnamefont {Sergeev}},\ and\ \bibinfo {author}
  {\bibfnamefont {M.}~\bibnamefont {Matheson}},\ }\bibfield  {title} {\bibinfo
  {title} {Exascale deep learning for scientific inverse problems},\
  }\href@noop {} {\bibfield  {journal} {\bibinfo  {journal} {arXiv e-prints}\ }
  (\bibinfo {year} {2019})},\ \Eprint {https://arxiv.org/abs/1909.11150}
  {arXiv:1909.11150 [cs.LG]} \BibitemShut {NoStop}%
\bibitem [{\citenamefont {Takeuchi}\ \emph {et~al.}(2003)\citenamefont
  {Takeuchi}, \citenamefont {Thamizhavel}, \citenamefont {Okubo}, \citenamefont
  {Yamada}, \citenamefont {Nakamura}, \citenamefont {Yamamoto}, \citenamefont
  {Inada}, \citenamefont {Sugiyama}, \citenamefont {Galatanu}, \citenamefont
  {Yamamoto}, \citenamefont {Kindo}, \citenamefont {Ebihara},\ and\
  \citenamefont {\ifmmode~\bar{O}\else \={O}\fi{}nuki}}]{Takeuchi-PRB-2003}%
  \BibitemOpen
  \bibfield  {author} {\bibinfo {author} {\bibfnamefont {T.}~\bibnamefont
  {Takeuchi}}, \bibinfo {author} {\bibfnamefont {A.}~\bibnamefont
  {Thamizhavel}}, \bibinfo {author} {\bibfnamefont {T.}~\bibnamefont {Okubo}},
  \bibinfo {author} {\bibfnamefont {M.}~\bibnamefont {Yamada}}, \bibinfo
  {author} {\bibfnamefont {N.}~\bibnamefont {Nakamura}}, \bibinfo {author}
  {\bibfnamefont {T.}~\bibnamefont {Yamamoto}}, \bibinfo {author}
  {\bibfnamefont {Y.}~\bibnamefont {Inada}}, \bibinfo {author} {\bibfnamefont
  {K.}~\bibnamefont {Sugiyama}}, \bibinfo {author} {\bibfnamefont
  {A.}~\bibnamefont {Galatanu}}, \bibinfo {author} {\bibfnamefont
  {E.}~\bibnamefont {Yamamoto}}, \bibinfo {author} {\bibfnamefont
  {K.}~\bibnamefont {Kindo}}, \bibinfo {author} {\bibfnamefont
  {T.}~\bibnamefont {Ebihara}},\ and\ \bibinfo {author} {\bibfnamefont
  {Y.}~\bibnamefont {\ifmmode~\bar{O}\else \={O}\fi{}nuki}},\ }\bibfield
  {title} {\bibinfo {title} {Anisotropic, thermal, and magnetic properties of
  {{CeAgSb$_2$}}: Explanation via a crystalline electric field scheme},\ }\href
  {https://doi.org/10.1103/PhysRevB.67.064403} {\bibfield  {journal} {\bibinfo
  {journal} {Phys. Rev. B}\ }\textbf {\bibinfo {volume} {67}},\ \bibinfo
  {pages} {064403} (\bibinfo {year} {2003})}\BibitemShut {NoStop}%
\bibitem [{\citenamefont {Edmonds}(1957)}]{edmondsAngularMomentumQuantum1957}%
  \BibitemOpen
  \bibfield  {author} {\bibinfo {author} {\bibfnamefont {A.~R.}\ \bibnamefont
  {Edmonds}},\ }\href@noop {} {\emph {\bibinfo {title} {Angular Momentum in
  Quantum Mechanics}}}\ (\bibinfo  {publisher} {{Princeton Univ. Press}},\
  \bibinfo {address} {{Princeton, N.J.}},\ \bibinfo {year} {1957})\BibitemShut
  {NoStop}%
\bibitem [{\citenamefont {Judd}(1963)}]{juddOperatorTechniquesAtomic1963}%
  \BibitemOpen
  \bibfield  {author} {\bibinfo {author} {\bibfnamefont {B.~R.}\ \bibnamefont
  {Judd}},\ }\href@noop {} {\emph {\bibinfo {title} {Operator Techniques in
  Atomic Spectroscopy}}},\ \bibinfo {edition} {1st}\ ed.\ (\bibinfo
  {publisher} {{McGraw-Hill}},\ \bibinfo {address} {{New York}},\ \bibinfo
  {year} {1963})\BibitemShut {NoStop}%
\bibitem [{\citenamefont
  {Buckmaster}(1962)}]{buckmasterTablesMatrixElements1962}%
  \BibitemOpen
  \bibfield  {author} {\bibinfo {author} {\bibfnamefont {H.~A.}\ \bibnamefont
  {Buckmaster}},\ }\bibfield  {title} {\bibinfo {title} {Tables of matrix
  elements for the operators},\ }\href {https://doi.org/10.1139/p62-171}
  {\bibfield  {journal} {\bibinfo  {journal} {Can. J. Phys.}\ }\textbf
  {\bibinfo {volume} {40}},\ \bibinfo {pages} {1670} (\bibinfo {year}
  {1962})}\BibitemShut {NoStop}%
\bibitem [{\citenamefont {Smith}\ and\ \citenamefont
  {Thornley}(1966)}]{smithUseOperatorEquivalents1966}%
  \BibitemOpen
  \bibfield  {author} {\bibinfo {author} {\bibfnamefont {D.}~\bibnamefont
  {Smith}}\ and\ \bibinfo {author} {\bibfnamefont {J.~H.~M.}\ \bibnamefont
  {Thornley}},\ }\bibfield  {title} {\bibinfo {title} {The use of 'operator
  equivalents'},\ }\href {https://doi.org/10.1088/0370-1328/89/4/301}
  {\bibfield  {journal} {\bibinfo  {journal} {Proc. Phys. Soc.}\ }\textbf
  {\bibinfo {volume} {89}},\ \bibinfo {pages} {779} (\bibinfo {year}
  {1966})}\BibitemShut {NoStop}%
\bibitem [{\citenamefont {Danielsen}\ and\ \citenamefont
  {Lindg{\aa}rd}(1972)}]{danielsenQuantumMechanicalOperator1972}%
  \BibitemOpen
  \bibfield  {author} {\bibinfo {author} {\bibfnamefont {O.}~\bibnamefont
  {Danielsen}}\ and\ \bibinfo {author} {\bibfnamefont {P.-A.}\ \bibnamefont
  {Lindg{\aa}rd}},\ }\href@noop {} {\emph {\bibinfo {title} {Quantum Mechanical
  Operator Equivalents Used in the Theory of Magnetism}}},\ \bibinfo {series}
  {Denmark. {{Forskningscenter}} Risoe. {{Risoe}}-r}\ No.\ \bibinfo {number}
  {259}\ (\bibinfo  {publisher} {{Ris\o{} National Laboratory}},\ \bibinfo
  {year} {1972})\BibitemShut {NoStop}%
\bibitem [{\citenamefont {Abragam}\ and\ \citenamefont
  {Bleaney}(1970)}]{abragamElectronParamagneticResonance1970}%
  \BibitemOpen
  \bibfield  {author} {\bibinfo {author} {\bibfnamefont {A.}~\bibnamefont
  {Abragam}}\ and\ \bibinfo {author} {\bibfnamefont {B.}~\bibnamefont
  {Bleaney}},\ }\href@noop {} {\emph {\bibinfo {title} {Electron Paramagnetic
  Resonance of Transition Ions}}}\ (\bibinfo  {publisher} {{Clarendon Press ;
  Oxford University Press}},\ \bibinfo {address} {{Oxford}},\ \bibinfo {year}
  {1970})\BibitemShut {NoStop}%
\bibitem [{\citenamefont {Mallat}(2008)}]{mallatWaveletTourSignal2008}%
  \BibitemOpen
  \bibfield  {author} {\bibinfo {author} {\bibfnamefont {S.}~\bibnamefont
  {Mallat}},\ }\href@noop {} {\emph {\bibinfo {title} {A {{Wavelet Tour}} of
  {{Signal Processing}}: {{The Sparse Way}}}}},\ \bibinfo {edition} {3rd}\ ed.\
  (\bibinfo  {publisher} {{Academic Press}},\ \bibinfo {address} {{Amsterdam ;
  Boston}},\ \bibinfo {year} {2008})\BibitemShut {NoStop}%
\bibitem [{\citenamefont {Lee}\ \emph {et~al.}(2019)\citenamefont {Lee},
  \citenamefont {Gommers}, \citenamefont {Wohlfahrt}, \citenamefont
  {Wasilewski}, \citenamefont {O'Leary}, \citenamefont {Nahrstaedt},
  \citenamefont {Hurtado}, \citenamefont {Sauv{\'e}}, \citenamefont {Arildsen},
  \citenamefont {Oliveira}, \citenamefont {Pelt}, \citenamefont {Agrawal},
  \citenamefont {{SylvainLan}}, \citenamefont {Pelletier}, \citenamefont
  {Brett}, \citenamefont {Yu}, \citenamefont {Choudhary}, \citenamefont
  {Tricoli}, \citenamefont {Craig}, \citenamefont {Ravindranathan},
  \citenamefont {Dan}, \citenamefont {{jakirkham}}, \citenamefont {Antonello},
  \citenamefont {Laszuk}, \citenamefont {Goertzen}, \citenamefont {Goldberg},
  \citenamefont {Reczey}, \citenamefont {{0-tree}}, \citenamefont {Smith},\
  and\ \citenamefont {{asnt}}}]{leePyWaveletsPywtPyWavelets2019}%
  \BibitemOpen
  \bibfield  {author} {\bibinfo {author} {\bibfnamefont {G.~R.}\ \bibnamefont
  {Lee}}, \bibinfo {author} {\bibfnamefont {R.}~\bibnamefont {Gommers}},
  \bibinfo {author} {\bibfnamefont {K.}~\bibnamefont {Wohlfahrt}}, \bibinfo
  {author} {\bibfnamefont {F.}~\bibnamefont {Wasilewski}}, \bibinfo {author}
  {\bibfnamefont {A.}~\bibnamefont {O'Leary}}, \bibinfo {author} {\bibfnamefont
  {H.}~\bibnamefont {Nahrstaedt}}, \bibinfo {author} {\bibfnamefont {D.~M.}\
  \bibnamefont {Hurtado}}, \bibinfo {author} {\bibfnamefont {A.}~\bibnamefont
  {Sauv{\'e}}}, \bibinfo {author} {\bibfnamefont {T.}~\bibnamefont {Arildsen}},
  \bibinfo {author} {\bibfnamefont {H.}~\bibnamefont {Oliveira}}, \bibinfo
  {author} {\bibfnamefont {D.~M.}\ \bibnamefont {Pelt}}, \bibinfo {author}
  {\bibfnamefont {A.}~\bibnamefont {Agrawal}}, \bibinfo {author} {\bibnamefont
  {{SylvainLan}}}, \bibinfo {author} {\bibfnamefont {M.}~\bibnamefont
  {Pelletier}}, \bibinfo {author} {\bibfnamefont {M.}~\bibnamefont {Brett}},
  \bibinfo {author} {\bibfnamefont {F.}~\bibnamefont {Yu}}, \bibinfo {author}
  {\bibfnamefont {S.}~\bibnamefont {Choudhary}}, \bibinfo {author}
  {\bibfnamefont {D.}~\bibnamefont {Tricoli}}, \bibinfo {author} {\bibfnamefont
  {L.~M.}\ \bibnamefont {Craig}}, \bibinfo {author} {\bibfnamefont
  {L.}~\bibnamefont {Ravindranathan}}, \bibinfo {author} {\bibfnamefont
  {J.}~\bibnamefont {Dan}}, \bibinfo {author} {\bibnamefont {{jakirkham}}},
  \bibinfo {author} {\bibfnamefont {J.}~\bibnamefont {Antonello}}, \bibinfo
  {author} {\bibfnamefont {D.}~\bibnamefont {Laszuk}}, \bibinfo {author}
  {\bibfnamefont {D.}~\bibnamefont {Goertzen}}, \bibinfo {author}
  {\bibfnamefont {C.}~\bibnamefont {Goldberg}}, \bibinfo {author}
  {\bibfnamefont {B.}~\bibnamefont {Reczey}}, \bibinfo {author} {\bibnamefont
  {{0-tree}}}, \bibinfo {author} {\bibfnamefont {A.}~\bibnamefont {Smith}},\
  and\ \bibinfo {author} {\bibnamefont {{asnt}}},\ }\href
  {https://doi.org/10.5281/zenodo.3510098} {\bibinfo {title}
  {{{PyWavelets}}/pywt: {{PyWavelets}} 1.1.1}},\ \bibinfo {howpublished}
  {Zenodo} (\bibinfo {year} {2019})\BibitemShut {NoStop}%
\bibitem [{\citenamefont {Lecun}\ \emph {et~al.}(1998)\citenamefont {Lecun},
  \citenamefont {Bottou}, \citenamefont {Bengio},\ and\ \citenamefont
  {Haffner}}]{Lecun98gradient-basedlearning}%
  \BibitemOpen
  \bibfield  {author} {\bibinfo {author} {\bibfnamefont {Y.}~\bibnamefont
  {Lecun}}, \bibinfo {author} {\bibfnamefont {L.}~\bibnamefont {Bottou}},
  \bibinfo {author} {\bibfnamefont {Y.}~\bibnamefont {Bengio}},\ and\ \bibinfo
  {author} {\bibfnamefont {P.}~\bibnamefont {Haffner}},\ }\bibfield  {title}
  {\bibinfo {title} {Gradient-based learning applied to document recognition},\
  }in\ \href@noop {} {\emph {\bibinfo {booktitle} {Proceedings of the
  {{IEEE}}}}}\ (\bibinfo {year} {1998})\ pp.\ \bibinfo {pages}
  {2278--2324}\BibitemShut {NoStop}%
\bibitem [{\citenamefont {Agarap}(2018)}]{DBLP:journals/corr/abs-1803-08375}%
  \BibitemOpen
  \bibfield  {author} {\bibinfo {author} {\bibfnamefont {A.~F.}\ \bibnamefont
  {Agarap}},\ }\bibfield  {title} {\bibinfo {title} {Deep learning using
  rectified linear units ({{ReLU}})},\ }\href@noop {} {\bibfield  {journal}
  {\bibinfo  {journal} {CoRR}\ } (\bibinfo {year} {2018})},\ \Eprint
  {https://arxiv.org/abs/1803.08375} {arXiv:1803.08375} \BibitemShut {NoStop}%
\bibitem [{\citenamefont {Chollet}\ \emph {et~al.}(2015)\citenamefont {Chollet}
  \emph {et~al.}}]{chollet2015keras}%
  \BibitemOpen
  \bibfield  {author} {\bibinfo {author} {\bibfnamefont {F.}~\bibnamefont
  {Chollet}} \emph {et~al.},\ }\href@noop {} {\bibinfo {title} {Keras}},\
  \bibinfo {howpublished} {https://keras.io} (\bibinfo {year}
  {2015})\BibitemShut {NoStop}%
\bibitem [{\citenamefont {Kingma}\ and\ \citenamefont
  {Ba}(2015)}]{DBLP:journals/corr/KingmaB14}%
  \BibitemOpen
  \bibfield  {author} {\bibinfo {author} {\bibfnamefont {D.~P.}\ \bibnamefont
  {Kingma}}\ and\ \bibinfo {author} {\bibfnamefont {J.}~\bibnamefont {Ba}},\
  }\bibfield  {title} {\bibinfo {title} {Adam: {{A}} method for stochastic
  optimization},\ }in\ \href@noop {} {\emph {\bibinfo {booktitle} {3rd
  International Conference on Learning Representations, {{ICLR}} 2015, San
  Diego, {{CA}}, {{USA}}, May 7-9, 2015, Conference Track Proceedings}}},\
  \bibinfo {editor} {edited by\ \bibinfo {editor} {\bibfnamefont
  {Y.}~\bibnamefont {Bengio}}\ and\ \bibinfo {editor} {\bibfnamefont
  {Y.}~\bibnamefont {LeCun}}}\ (\bibinfo {year} {2015})\BibitemShut {NoStop}%
\bibitem [{\citenamefont {Orth}\ and\ \citenamefont
  {Berthusen}(2020)}]{orth_berthusen_2020}%
  \BibitemOpen
  \bibfield  {author} {\bibinfo {author} {\bibfnamefont {P.~P.}\ \bibnamefont
  {Orth}}\ and\ \bibinfo {author} {\bibfnamefont {N.}~\bibnamefont
  {Berthusen}},\ }\href {https://doi.org/10.6084/m9.figshare.13321985.v2}
  {\bibinfo {title} {Using convolutional neural networks to extract crystal
  field parameters from thermodynamic data}} (\bibinfo {year}
  {2020})\BibitemShut {NoStop}%
\bibitem [{\citenamefont {Jobiliong}\ \emph {et~al.}(2005)\citenamefont
  {Jobiliong}, \citenamefont {Brooks}, \citenamefont {Choi}, \citenamefont
  {Lee},\ and\ \citenamefont {Fisk}}]{Jobiliong-PRB-2005}%
  \BibitemOpen
  \bibfield  {author} {\bibinfo {author} {\bibfnamefont {E.}~\bibnamefont
  {Jobiliong}}, \bibinfo {author} {\bibfnamefont {J.~S.}\ \bibnamefont
  {Brooks}}, \bibinfo {author} {\bibfnamefont {E.~S.}\ \bibnamefont {Choi}},
  \bibinfo {author} {\bibfnamefont {H.}~\bibnamefont {Lee}},\ and\ \bibinfo
  {author} {\bibfnamefont {Z.}~\bibnamefont {Fisk}},\ }\bibfield  {title}
  {\bibinfo {title} {Magnetization and electrical-transport investigation of
  the dense kondo system $\mathrm{Ce}\mathrm{Ag}{\mathrm{sb}}_{2}$},\ }\href
  {https://doi.org/10.1103/PhysRevB.72.104428} {\bibfield  {journal} {\bibinfo
  {journal} {Phys. Rev. B}\ }\textbf {\bibinfo {volume} {72}},\ \bibinfo
  {pages} {104428} (\bibinfo {year} {2005})}\BibitemShut {NoStop}%
\bibitem [{\citenamefont {Princep}\ \emph {et~al.}(2013)\citenamefont
  {Princep}, \citenamefont {Prabhakaran}, \citenamefont {Boothroyd},\ and\
  \citenamefont {Adroja}}]{Princep-PRB-2013}%
  \BibitemOpen
  \bibfield  {author} {\bibinfo {author} {\bibfnamefont {A.~J.}\ \bibnamefont
  {Princep}}, \bibinfo {author} {\bibfnamefont {D.}~\bibnamefont
  {Prabhakaran}}, \bibinfo {author} {\bibfnamefont {A.~T.}\ \bibnamefont
  {Boothroyd}},\ and\ \bibinfo {author} {\bibfnamefont {D.~T.}\ \bibnamefont
  {Adroja}},\ }\bibfield  {title} {\bibinfo {title} {Crystal-field states of
  pr${}^{3+}$ in the candidate quantum spin ice
  pr${}_{2}$sn${}_{2}$o${}_{7}$},\ }\href
  {https://doi.org/10.1103/PhysRevB.88.104421} {\bibfield  {journal} {\bibinfo
  {journal} {Phys. Rev. B}\ }\textbf {\bibinfo {volume} {88}},\ \bibinfo
  {pages} {104421} (\bibinfo {year} {2013})}\BibitemShut {NoStop}%
\bibitem [{\citenamefont {Takeuchi}\ \emph {et~al.}(2001)\citenamefont
  {Takeuchi}, \citenamefont {Inoue}, \citenamefont {Sugiyama}, \citenamefont
  {Aoki}, \citenamefont {Tokiwa}, \citenamefont {Haga}, \citenamefont {Kindo},\
  and\ \citenamefont {Ōnuki}}]{Takeuchi-CeRhIn5-2001}%
  \BibitemOpen
  \bibfield  {author} {\bibinfo {author} {\bibfnamefont {T.}~\bibnamefont
  {Takeuchi}}, \bibinfo {author} {\bibfnamefont {T.}~\bibnamefont {Inoue}},
  \bibinfo {author} {\bibfnamefont {K.}~\bibnamefont {Sugiyama}}, \bibinfo
  {author} {\bibfnamefont {D.}~\bibnamefont {Aoki}}, \bibinfo {author}
  {\bibfnamefont {Y.}~\bibnamefont {Tokiwa}}, \bibinfo {author} {\bibfnamefont
  {Y.}~\bibnamefont {Haga}}, \bibinfo {author} {\bibfnamefont {K.}~\bibnamefont
  {Kindo}},\ and\ \bibinfo {author} {\bibfnamefont {Y.}~\bibnamefont
  {Ōnuki}},\ }\bibfield  {title} {\bibinfo {title} {Magnetic and thermal
  properties of ceirin5 and cerhin5},\ }\href
  {https://doi.org/10.1143/JPSJ.70.877} {\bibfield  {journal} {\bibinfo
  {journal} {Journal of the Physical Society of Japan}\ }\textbf {\bibinfo
  {volume} {70}},\ \bibinfo {pages} {877} (\bibinfo {year} {2001})},\ \Eprint
  {https://arxiv.org/abs/https://doi.org/10.1143/JPSJ.70.877}
  {https://doi.org/10.1143/JPSJ.70.877} \BibitemShut {NoStop}%
\bibitem [{\citenamefont {Araki}\ \emph {et~al.}(2003)\citenamefont {Araki},
  \citenamefont {Metoki}, \citenamefont {Galatanu}, \citenamefont {Yamamoto},
  \citenamefont {Thamizhavel},\ and\ \citenamefont {\ifmmode~\bar{O}\else
  \={O}\fi{}nuki}}]{Araki-PRB-2003}%
  \BibitemOpen
  \bibfield  {author} {\bibinfo {author} {\bibfnamefont {S.}~\bibnamefont
  {Araki}}, \bibinfo {author} {\bibfnamefont {N.}~\bibnamefont {Metoki}},
  \bibinfo {author} {\bibfnamefont {A.}~\bibnamefont {Galatanu}}, \bibinfo
  {author} {\bibfnamefont {E.}~\bibnamefont {Yamamoto}}, \bibinfo {author}
  {\bibfnamefont {A.}~\bibnamefont {Thamizhavel}},\ and\ \bibinfo {author}
  {\bibfnamefont {Y.}~\bibnamefont {\ifmmode~\bar{O}\else \={O}\fi{}nuki}},\
  }\bibfield  {title} {\bibinfo {title} {Crystal structure, magnetic ordering,
  and magnetic excitation in the $4f$-localized ferromagnet
  ${\mathrm{ceagsb}}_{2}$},\ }\href
  {https://doi.org/10.1103/PhysRevB.68.024408} {\bibfield  {journal} {\bibinfo
  {journal} {Phys. Rev. B}\ }\textbf {\bibinfo {volume} {68}},\ \bibinfo
  {pages} {024408} (\bibinfo {year} {2003})}\BibitemShut {NoStop}%
\bibitem [{\citenamefont {Hafner}\ \emph {et~al.}(2019)\citenamefont {Hafner},
  \citenamefont {Rai}, \citenamefont {Banda}, \citenamefont {Kliemt},
  \citenamefont {Krellner}, \citenamefont {Sichelschmidt}, \citenamefont
  {Morosan}, \citenamefont {Geibel},\ and\ \citenamefont
  {Brando}}]{Hafner-PRB-2019}%
  \BibitemOpen
  \bibfield  {author} {\bibinfo {author} {\bibfnamefont {D.}~\bibnamefont
  {Hafner}}, \bibinfo {author} {\bibfnamefont {B.~K.}\ \bibnamefont {Rai}},
  \bibinfo {author} {\bibfnamefont {J.}~\bibnamefont {Banda}}, \bibinfo
  {author} {\bibfnamefont {K.}~\bibnamefont {Kliemt}}, \bibinfo {author}
  {\bibfnamefont {C.}~\bibnamefont {Krellner}}, \bibinfo {author}
  {\bibfnamefont {J.}~\bibnamefont {Sichelschmidt}}, \bibinfo {author}
  {\bibfnamefont {E.}~\bibnamefont {Morosan}}, \bibinfo {author} {\bibfnamefont
  {C.}~\bibnamefont {Geibel}},\ and\ \bibinfo {author} {\bibfnamefont
  {M.}~\bibnamefont {Brando}},\ }\bibfield  {title} {\bibinfo {title}
  {Kondo-lattice ferromagnets and their peculiar order along the magnetically
  hard axis determined by the crystalline electric field},\ }\href
  {https://doi.org/10.1103/PhysRevB.99.201109} {\bibfield  {journal} {\bibinfo
  {journal} {Phys. Rev. B}\ }\textbf {\bibinfo {volume} {99}},\ \bibinfo
  {pages} {201109} (\bibinfo {year} {2019})}\BibitemShut {NoStop}%
\bibitem [{\citenamefont {Wang}(1971)}]{Wang-Phys_Lett_A-1971}%
  \BibitemOpen
  \bibfield  {author} {\bibinfo {author} {\bibfnamefont {Y.-L.}\ \bibnamefont
  {Wang}},\ }\bibfield  {title} {\bibinfo {title} {Crystal-field effects of
  paramagnetic curie temperature},\ }\href
  {https://doi.org/https://doi.org/10.1016/0375-9601(71)90750-X} {\bibfield
  {journal} {\bibinfo  {journal} {Physics Letters A}\ }\textbf {\bibinfo
  {volume} {35}},\ \bibinfo {pages} {383} (\bibinfo {year} {1971})}\BibitemShut
  {NoStop}%
\bibitem [{\citenamefont
  {Johnston}(2015)}]{Johnston-Unified_molecular_field-PRB-2015}%
  \BibitemOpen
  \bibfield  {author} {\bibinfo {author} {\bibfnamefont {D.~C.}\ \bibnamefont
  {Johnston}},\ }\bibfield  {title} {\bibinfo {title} {Unified molecular field
  theory for collinear and noncollinear heisenberg antiferromagnets},\ }\href
  {https://doi.org/10.1103/PhysRevB.91.064427} {\bibfield  {journal} {\bibinfo
  {journal} {Phys. Rev. B}\ }\textbf {\bibinfo {volume} {91}},\ \bibinfo
  {pages} {064427} (\bibinfo {year} {2015})}\BibitemShut {NoStop}%
\bibitem [{\citenamefont {Goodfellow}\ \emph {et~al.}(2014)\citenamefont
  {Goodfellow}, \citenamefont {Pouget-Abadie}, \citenamefont {Mirza},
  \citenamefont {Xu}, \citenamefont {Warde-Farley}, \citenamefont {Ozair},
  \citenamefont {Courville},\ and\ \citenamefont
  {Bengio}}]{goodfellow2014generative}%
  \BibitemOpen
  \bibfield  {author} {\bibinfo {author} {\bibfnamefont {I.~J.}\ \bibnamefont
  {Goodfellow}}, \bibinfo {author} {\bibfnamefont {J.}~\bibnamefont
  {Pouget-Abadie}}, \bibinfo {author} {\bibfnamefont {M.}~\bibnamefont
  {Mirza}}, \bibinfo {author} {\bibfnamefont {B.}~\bibnamefont {Xu}}, \bibinfo
  {author} {\bibfnamefont {D.}~\bibnamefont {Warde-Farley}}, \bibinfo {author}
  {\bibfnamefont {S.}~\bibnamefont {Ozair}}, \bibinfo {author} {\bibfnamefont
  {A.}~\bibnamefont {Courville}},\ and\ \bibinfo {author} {\bibfnamefont
  {Y.}~\bibnamefont {Bengio}},\ }\href@noop {} {\bibinfo {title} {Generative
  adversarial networks}} (\bibinfo {year} {2014}),\ \Eprint
  {https://arxiv.org/abs/1406.2661} {arXiv:1406.2661 [stat.ML]} \BibitemShut
  {NoStop}%
\end{thebibliography}
%

\end{document}